\documentclass[a4paper, 12pt]{article}
\usepackage[margin=0.7in]{geometry}  
\usepackage[T1]{fontenc}
\usepackage[utf8]{inputenc}
\usepackage{microtype}
\usepackage{graphicx} 
\usepackage{amsfonts}       
\usepackage{breqn}
\usepackage{url}           
\usepackage{listings}
\usepackage{xcolor}
\usepackage{caption}
\usepackage{subcaption}
\usepackage{multirow} 
\usepackage{cite}
\usepackage[colorlinks=true, linkcolor=blue, citecolor=blue, urlcolor=blue]{hyperref}
\usepackage{float}
\usepackage[stable]{footmisc}
\usepackage{tablefootnote}
\usepackage{amsmath}
\usepackage{array}
\usepackage{setspace}
\usepackage{booktabs}
\usepackage{tabularx}
\usepackage{booktabs}

\newcommand{\beq}{\begin{eqnarray}}
\newcommand{\eeq}{\end{eqnarray}}
\newcommand{\bpmatrix}{\begin{pmatrix}}
\newcommand{\epmatrix}{\end{pmatrix}}
\newcommand{\ba}{\begin{array}}
\newcommand{\ea}{\end{array}}

\def\bea{\begin{eqnarray}}   
\def\eea{\end{eqnarray}}

\title{Confronting the broken phase N2HDM with Higgs Data}

\author{
  Maien Binjonaid \\
  Department of Physics and Astronomy\\
  King Saud University\\
  Riyadh, Saudi Arabia \\
  \texttt{maien@ksu.edu.sa} \\
}

\begin{document}

\maketitle

\doublespacing

\begin{abstract}
The broken phase of the Next-to two-Higgs-doublet model (N2HDM) constitutes an archetype of extended Higgs sectors. In the presence of a softly-broken $\mathrm{Z}_2$ symmetry throughout the scalar and Yukawa sectors, as the additional gauge singlet field does not interact with fermions, the model admits four variants of Yukawa interactions between the doublets and Standard Model fermions. We confront each type with experimental Higgs data, especially from CMS and ATLAS detectors at the LHC. Interfacing the models with the the state-of-the-art package $\mathtt{HiggsTools}$, we perform a statistical $\chi^2$ analysis to determine the best-fit points and exclusion limits at the $95\%$ and $68\%$ C.L., and identify SM-like Higgs measurements that affect each type the most. We further analyze the exclusion bounds on the additional Higgs bosons at the $95\%$ C.L., paying special attention to searches of hypothetical non-SM Higgs resonances decaying into a pair of bosons or fermions. We show regions where the additional Higgs bosons do not satisfy the narrow-width approximation utilized in most experimental searches.

\end{abstract}

\section{Introduction} \label{intro}
The next-to two-higgs-doublet model (N2HDM) is a well-motivated Beyond the Standard Model (BSM) extension, providing a plethora of new possibilities for direct and indirect experimental searches of new physics (see \cite{Chen:2013jvg, Muhlleitner:2016mzt} and references therein). This is due to the generic structure of its Higgs sector, which contains an additional real singlet compared to the 2HDM. Generally, there are different aspects that motivate the study of doublet and singlet extensions of the Standard Model (SM) \cite{Ivanov:2017dad}. Among these are the continuing efforts to understand the nature of electroweak symmetry breaking (EWSB), and searches for deviations from SM due to the possibility of the presence of additional Higgs doublets and/or singlets. Given that the Large Hadron Collider (LHC) is consistently examining the properties of the discovered scalar and searching for additional scalars predicted by such extensions \cite{CMS:2022dwd, ATLAS:2022vkf}, it is vital to understand to what extent those deviations are allowed and to confront BSM extensions with experimental Higgs data. 

In fact, several aspects of the extended SM with singlets and doublets were considered in the literature, including collider phenomenology, dark matter, and cosmology \cite{Barger:2008jx,Guo:2010hq, Biswas:2011td, Costa:2015llh, GAMBIT:2017gge, Robens:2019kga, aali:2020tgr, Basak:2021rzq, Coito:2021rdb, Ellis:2022lft, Drozd:2014yla,Baum:2018ekm, MoortgatPick:2023siz,Bhattacharya:2023gfi,Paasch:2023pjq, Darvishi:2022wnd}. The N2HDM can be seen as a baseline model that captures a range of phenomenological effects that arise from adding new doublets and singlets to the Higgs potential. It is the extension of the CP-conserving 2HDM by an additional real singlet\footnote{The CP-violating variant was investigated in \cite{Muhlleitner:2021cci}}, where two $\mathrm{Z}_2$ symmetries are imposed to eliminate flavor-changing neutral currents (FCNC) at tree-level, and possibly providing a stable dark matter (DM) candidate. Its vacuum structure is more involved than CP-conserving 2HDMs or singlet extensions, since a vacuum expectation value (VEV) can be acquired by two Higgs doublets and the singlet (the broken phase), only two Higgs doublets (the dark singlet phase), only one Higgs doublet and the singlet (the dark doublet phase), or only one Higgs doublet (the fully dark phase). In the first case, no DM candidate is present, while the other cases can provide a singlet DM, a doublet DM, or two DM candidates (see \cite{Engeln:2020fld} for details and references to earlier work along this direction). 

As the two doublets can interact with SM fermions, and in the presence of a softly-broken $\mathrm{Z}_2$ symmetry preventing FCNC, the Yukawa sector inherits the four different types associated with the 2HDM \cite{i-1,i-2,ii-2,xy-1,xy-2,xy-3,xy-4,xy-5}: Type 1, Type 2, Type X (Lepton-Specific), and Type Y (Flipped). Other more generic scenarios for Yukawa interactions, without $\mathrm{Z}_2$ symmetry, have been considered in 2HDM \cite{Pich:2009sp, Branco:2011iw, Grzadkowski:2018ohf}. Moreover, in N2HDM, both the CP-odd state $A$ and the charged Higgs pair $H^{\pm}$ have the same structure as in 2HDM. However, the constraints specific to N2HDM may have some indirect effects on such states.

From the theoretical side, the renormalization of the full model was carried out in \cite{Krause:2017mal}, where it was shown that the effects of corrections can be sizable. The impact of electroweak corrections was thoroughly analyzed in \cite{Krause:2019oar} and implemented in \cite{Krause:2019qwe}. Cosmological aspects of the Type 2 model were investigated in \cite{Biekotter:2021ysx}, demonstrating cases where electroweak symmetry is not restored. A comprehensive analysis of vacuum instabilities was provided in \cite{Ferreira:2019iqb}. Additionally, the naturalness of the model was considered in \cite{Arhrib:2024itt}. 

From the phenomenological and experimental sides, an analysis of Types 1 and 2 in an approximated version of the model was performed in \cite{Chen:2013jvg} with $H_2$ considered SM-like, while a systematic analysis of the same types, focusing on wrong-sign regions and the singlet component of the SM-like Higgs boson (which could be any of the CP-even Higgs bosons), was conducted in \cite{Muhlleitner:2016mzt}. The CMS collaboration searched for resonant pair production of Higgs bosons in the $b\bar{b}ZZ$ final state, and provided an interpretation of the results specific to the N2HDM scenario \cite{CMS:2020jeo}. A dedicated investigation of di-Higgs production in the 4-photon final state was carried out in \cite{Arhrib:2018qmw} assuming Type 1, while a comprehensive analysis of the limits on di-Higgs production was presented in \cite{Abouabid:2021yvw} for Types 1 and 2, with careful consideration for the distinction between resonant and non-resonant regions. The accommodation of an additional 96 GeV Higgs boson was considered in \cite{Biekotter:2019kde, Biekotter:2020cjs, Biekotter:2022abc}. The prospects of the model for the future electron-positron collider were presented in \cite{Azevedo:2018llq}. The model is implemented in the public tool $\mathtt{ScannerS}$ \cite{Muhlleitner:2020wwk}, which facilitates sophisticated phenomenological studies.

Given the continuing interest and interchanging efforts from both the phenomenology and the experiment sides, especially with the advent of several new results from LHC SM-like Higgs measurements and searches for additional Higgs bosons, our aim in this paper is expand on previous works and confront all types of N2HDM with the latest Higgs data available in the public code $\mathtt{HiggsTools}$ (HT). Specifically, we interface the model with HT, carry out an up-to-date statistical $\chi^2$ analysis, obtaining exclusion limits on the SM-like Higgs boson with $95\%$ and $68\%$ Confidence Level (C.L.), as well as $95\%$ C.L. exclusion limits on the additional Higgs bosons, paying special attention to the pair production of bosons through a heavy scalar resonance. 

The paper is organized as follows. In Section 2, we describe the theoretical aspects of the Higgs sector. In Section 3, we present the analysis scheme and the constraints taken into account. In Section 4, we provide the results and discuss them in the context of recent LHC searches and measurements. Finally, we conclude in Section 5.

\section{The Higgs Sector}
\label{sec:core1}

In terms of the two $SU(2)_L$ Higgs doublets $\Phi_1$ and $\Phi_2$ and the real singlet field $\Phi_S$, the Higgs sector of the CP-conserving N2HDM is described by the following scalar potential:
\begin{align}
V_{\text{N2HDM}} &= V_{\text{2HDM}} + V_{\text{singlet}}, 
\end{align}
where,
\begin{align}
V_{\text{2HDM}} &= m_{11}^2 |\Phi_1|^2 + m_{22}^2 |\Phi_2|^2 - m_{12}^2 (\Phi_1^\dagger \Phi_2 + \text{h.c.}) \nonumber \\
&\quad + \frac{\lambda_1}{2} (\Phi_1^\dagger \Phi_1)^2 + \frac{\lambda_2}{2} (\Phi_2^\dagger \Phi_2)^2 \nonumber \\
&\quad + \lambda_3 (\Phi_1^\dagger \Phi_1)(\Phi_2^\dagger \Phi_2) + \lambda_4 (\Phi_1^\dagger \Phi_2)(\Phi_2^\dagger \Phi_1) \nonumber \\
&\quad + \frac{\lambda_5}{2} [(\Phi_1^\dagger \Phi_2)^2 + \text{h.c.}],
\end{align}
and:
\begin{align}
V_{\text{singlet}} &= \frac{1}{2} m_S^2 \Phi_S^2 + \frac{\lambda_6}{8} \Phi_S^4 + \frac{\lambda_7}{2} (\Phi_1^\dagger \Phi_1) \Phi_S^2 + \frac{\lambda_8}{2} (\Phi_2^\dagger \Phi_2) \Phi_S^2.
\end{align}

All parameters are assumed to be real:
\begin{itemize}
    \item $m_{11}^2$, $m_{22}^2$, $m_S^2$: Mass-squared parameters for $\Phi_1$, $\Phi_2$, and $\Phi_S$.
    \item $m_{12}^2$: Soft-breaking mass-squared parameter.
    \item $\lambda_1$-$\lambda_8$: Quartic couplings.
\end{itemize}

The structure of the potential is dictated not only by SM symmetries, but also by two additional discrete symmetries $\mathrm{Z}_2$ ($\Phi_{1,S}:$ even, $\Phi_2:$ odd) and $\mathrm{Z}_2^{'}$ ($\Phi_{1,2}:$ even, $\Phi_S:$ odd). The first symmetry is softly broken by the $m_{12}^2$ term, similar to the 2HDM case, while the other one is spontaneously broken once the singlet field acquires a VEV. This structure forbids cubic terms in the potential (see \cite{Muhlleitner:2016mzt} for more details).

The broken phase of the N2HDM is defined as the case where all fields obtain VEVs, 
\begin{align}
\left<\Phi_1\right>=\begin{pmatrix}
0 \\ \frac{v_1}{\sqrt{2}}
\end{pmatrix},\qquad \left<\Phi_2\right>=\begin{pmatrix}
0 \\ \frac{v_2}{\sqrt{2}}
\end{pmatrix},\qquad\left<\Phi_S\right>=v_s\,,
\label{eq:n2hdm}
\end{align}
where the factor $\frac{1}{\sqrt{2}}$ sets the convention $v = \sqrt{v_1 + v_2} = 246.22$ GeV for the electroweak VEV. At its minimum, the potential takes the form:
\begin{align}
V &= \frac{m_{11}^2 v_1^2}{2} + \frac{m_{22}^2 v_2^2}{2} - m_{12}^2 v_1 v_2 + \frac{\lambda_1 v_1^4}{8} + \frac{\lambda_2 v_2^4}{8} + \frac{\lambda_{345} v_1^2 v_2^2}{4} \notag \\
&\quad + \frac{m_S^2 v_S^2}{2} + \frac{\lambda_6 v_S^4}{8} + \frac{\lambda_7 v_1^2 v_S^2}{4} + \frac{\lambda_8 v_2^2 v_S^2}{4},
\end{align}
which is minimized by:
\begin{equation}
    \bigg(\frac{\partial V}{\partial v_1}\bigg|_{\text{min}} , \frac{\partial V}{\partial v_2}\bigg|_{\text{min}} , \frac{\partial V}{\partial v_S}\bigg|_{\text{min}} \bigg) = 0,
\end{equation}
resulting in three conditions:
\begin{align}
  m_{11}^2 &= m_{12}^2\frac{v_2}{v_1} - \frac{1}{2}(v_1^2\lambda_1 + v_2^2\lambda_{345} + v_S^2\lambda_7) \\
  m_{22}^2 &= m_{12}^2\frac{v_1}{v_2} - \frac{1}{2}(v_1^2\lambda_{345} + v_2^2\lambda_2 + v_S^2\lambda_8) \\
  m_S^2 &= -\frac{1}{2}(v_1^2\lambda_7 + v_2^2\lambda_8 + v_S^2\lambda_6),
\end{align}
where $\lambda_{345} = \lambda_3 + \lambda_4 + \lambda_5$. 

The fields in Eq. 2 and 3, can be parametrized by expanding their neutral components around the VEVs: 
\begin{equation}
    \phi_i^0 = \gamma_i(v_i + \rho_i + i \sigma_i), 
\end{equation}
where $i=1, 2, s$ with $\gamma_{1,2} = \frac{1}{\sqrt{2}}$, $\gamma_s =1$, and $\sigma_s =0$. Focusing on the CP-even neutral components we have:
\begin{align}
  \mathcal{R}[\phi_1^0] &= \frac{v_1 + \rho_1}{\sqrt{2}} \\
  \mathcal{R}[\phi_2^0] &= \frac{v_2 + \rho_2}{\sqrt{2}} \\
  \phi_S &= v_S + \rho_S
\end{align}
In the $\{\rho_1, \rho_2, \rho_{s} \}$ basis, the elements of the mass-squared matrix $\mathcal{M}^2_\rho$ can be derived by:
\begin{equation}
  M^2_{ij} = \left.\frac{\partial^2 V}{\partial \rho_i \partial \rho_j}\right|_{\rho_k=0}
\end{equation}
Substituting $v_1 = v\cos\beta$ and $v_2 = v\sin\beta$, where $\tan{\beta}= \frac{v_2}{v_1}$, and using the minimization conditions, the matrix elements of $\mathcal{M}^2_\rho$ are:
\begin{align}
  M^2_{11} &= \lambda_1 v^2\cos^2\beta + m_{12}^2\tan\beta \\
  M^2_{12} &= \lambda_{345} v^2\cos\beta\sin\beta - m_{12}^2 \\
  M^2_{13} &= \lambda_7 vv_S\cos\beta \\
  M^2_{22} &= \lambda_2 v^2\sin^2\beta + \frac{m_{12}^2}{\tan\beta} \\
  M^2_{23} &= \lambda_8 vv_S\sin\beta \\
  M^2_{33} &= \lambda_6 v_S^2
\end{align}

One can exploit the properties of this self-adjoint matrix to set an upper bound on the lightest eigenvalue ($\Lambda_{\text{min}}$) of its diagonal form $\mathcal{M}^2_H$ (corresponding to mass eigenstates squared: $m_{H_i}^2$). In particular, rotating the top-left $2 \times 2$ part of $\mathcal{M}^2_\rho$ by a unitary matrix defined in terms of the angle $\beta$, the upper bound is:
\begin{equation}
    \Lambda_{\min} \leq \min{(A, B)},
\end{equation}
where A and B are the diagonal elements of the rotated $2 \times 2$ submatrix:
\begin{align}
A &= v^2 \left( \lambda_1 \cos^4{\beta} + 2\, \lambda_{345} \cos^2{\beta} \sin^2{\beta} + \lambda_2 \sin^4{\beta} \right) \\
B &= \frac{1}{2} m_{12}^2 \left( 3 - \cos{2\beta} \right) \cot{\beta} + \sin^2{\beta} \left( v^2 \left( \lambda_1 + \lambda_2 - 2\, \lambda_{345} \right) \cos^2{\beta} + m_{12}^2 \tan{\beta} \right)
\end{align}

This sets an upper limit on the lightest CP-even Higgs boson in the model (in this paper, we consider this state as the SM-like Higgs boson). 
Formally, $\mathcal{M}^2_\rho$ can be diagonalized using its eigenvalues and eigenvectors, taking us from gauge eigenstates to mass eigenstates ($H_1, H_2, H_3$). Given the excessively long expressions, it is more convenient to define an orthogonal rotation matrix $R(\alpha_1, \alpha_2, \alpha_3)$ whose columns are related to the eigenvectors of $\mathcal{M}^2_\rho$. Different parameterizations are possible, and the one used in \cite{Muhlleitner:2016mzt} is:

\begin{equation}
R = \begin{pmatrix} 
c_{\alpha_1}c_{\alpha_2} & s_{\alpha_1}c_{\alpha_2} & s_{\alpha_2} \\
-c_{\alpha_1}s_{\alpha_2}s_{\alpha_3} - s_{\alpha_1}c_{\alpha_3} & c_{\alpha_1}c_{\alpha_3} - s_{\alpha_1}s_{\alpha_2}s_{\alpha_3} & c_{\alpha_2}s_{\alpha_3} \\
-c_{\alpha_1}s_{\alpha_2}c_{\alpha_3} + s_{\alpha_1}s_{\alpha_3} & -c_{\alpha_1}s_{\alpha_3} - s_{\alpha_1}s_{\alpha_2}c_{\alpha_3} & c_{\alpha_2}c_{\alpha_3}
\end{pmatrix},
\end{equation}
with shorthand notation: $c_\theta \equiv \cos{(\theta)}$ and $s_\theta \equiv \sin{(\theta)}$, and the mixing angles reside between $(-\frac{\pi}{2},\frac{\pi}{2})$. Using $R$, we obtain,

\begin{equation}
    \mathcal{M}^2_{H} = R \mathcal{M}^2_{\rho} R^T,
\end{equation}
where $\mathcal{M}^2_H$ is diagonal and can be arranged to have the ordering: $m_{H_1}^2 < m_{H_2}^2 < m_{H_3}^2$. Any of these states can be the SM-like, as was considered in \cite{Engeln:2020fld}; however, in this paper we consider $H_1$ to be SM-like.

In N2HDM, the CP-even Higgs mass eignestates \( H_i \) are mixtures of the gauge eigenstates:
\begin{equation} \label{Hrho}
    H_i = \sum_{j=1}^{3} R_{ij} \rho_j,
\end{equation}
where $i,j=1,2,3$. This introduces a singlet component $|R_{i3}|^2$ that features this extension. 

Furthermore, the coupling of \( H_i \) to the gauge bosons \( V = W, Z \) can be derived from the kinetic part of the Lagrangian:
\begin{equation} \label{L_k}
\mathcal{L}_{\text{kin}} = (D_\mu \Phi_1)^\dagger (D^\mu \Phi_1) + (D_\mu \Phi_2)^\dagger (D^\mu \Phi_2)
\end{equation}
wher $D_\mu$ is the covariant derivative. The gauge singlet does not couple directly to gauge bosons; hence its kinetic term is not included. After Electroweak symmetry breaking, field expansions, and rotations from gauge eigenstates to CP-even mass eigenstates using:
\begin{equation}
 \rho_j = \sum_{i=1}^{3} R_{i j} H_i,   
\end{equation}
where the sum is over $i$, and only $j=1,2$ contribute to this coupling based on Eq.\ref{L_k}. One can directly derive the result:
\begin{equation}
   \mathcal{L}_{H_i VV}^{\text{N2HDM}} = c(H_iVV) \mathcal{L}_{hVV}^{\text{SM}}, 
\end{equation}
where the effective coupling is:
\begin{equation}
 c(H_i VV) = \frac{v_1}{v} R_{i1} + \frac{v_2}{v} R_{i2} = \cos\beta\, R_{i1} + \sin\beta\, R_{i2}.   
\end{equation}
For \( H_1 \), and substituting for the elements of \( R \), the coupling becomes:
\begin{equation}
    c(H_1 VV) = \cos\alpha_2\, \cos(\beta - \alpha_1).
\end{equation}

This shows that the deviation from the SM is driven by the mixing angles. In the limit where $\alpha_{2} \rightarrow 0$ and $\alpha_1 \rightarrow \alpha + \frac{\pi}{2}$ we recover the 2HDM coupling, while in the limit where \( \alpha_2 \to 0 \) and \( \alpha_1 \to \beta \), we recover the SM coupling (the alignment limit). More details on the alignment limit in 2HDMs can be found in \cite{Gunion:2002zf, Craig:2013hca, Carena:2013ooa}. 

The Yukawa Lagrangian of the N2HDM, before electroweak symmetry breaking, can be written similarly to the 2HDM, with the singlet field being inert with respect to fermions. Imposing a softly broken $\mathrm{Z}_2$ symmetry on the scalar and Yukawa sectors ensures the absence of flavour-changing neutral currents and leads to four possible types of Yukawa assignments:
\begin{itemize}
    \item Type 1 (T1): All fermions are exclusively coupled to $\Phi_2$.
    \item Type 2 (T2): Up-type quarks couple to $\Phi_2$, while down-type quarks and leptons couple to $\Phi_1$.
    \item Type X (TX): Quarks couple to $\Phi_2$, leptons couple to $\Phi_1$.
    \item Type Y (TY): Up-type quarks and leptons couple to $\Phi_2$, down-type quarks couple to $\Phi_1$.
\end{itemize}

As an example, consider the top Yukawa coupling in Type 1. After electroweak symmetry breaking and rotating to mass eigenstates $H_i$, the effective coupling of the lightest CP-even Higgs boson $H_1$ to the top quark relative to the SM can be expressed as:
\begin{equation}
  C(H_1 t\bar{t}) = \frac{R_{12}}{\sin\beta}.  
\end{equation}
In the SM limit, we have $R_{12} \to \sin\beta$, such that $C(H_1 t\bar{t}) \to 1$. The full list of effective couplings is presented in the appendix.

Finally, one can express the quartic couplings in the scalar potential in terms of physical masses, mixing angles, and VEVs:

\begin{equation}
\begin{array}{rcl}
\lambda_1 &=& \frac{1}{v^2 c_\beta^2} \left( -\frac{m_{12}^2 s_\beta}{c_\beta } + m_{H_1}^2 R_{11}^2 + m_{H_2}^2 R_{21}^2 + m_{H_3}^2 R_{31}^2 \right) \\[2mm]
\lambda_2 &=& \frac{1}{v^2 s_\beta^2} \left( -\frac{m_{12}^2 c_\beta}{ s_\beta} + m_{H_1}^2 R_{12}^2 + m_{H_2}^2 R_{22}^2 + m_{H_3}^2 R_{32}^2 \right) \\[2mm]
\lambda_3 &=& \frac{1}{v^2} \left( -\frac{m_{12}^2}{c_\beta s_\beta} + \frac{1}{s_\beta c_\beta}(m_{H_1}^2 R_{11}R_{12} + m_{H_2}^2 R_{21}R_{22} + m_{H_3}^2 R_{31}R_{32}) + 2m_{H^\pm}^2 \right) \\[2mm]
\lambda_4 &=& \frac{1}{v^2} \left( \frac{m_{12}^2}{c_\beta s_\beta} + m_A^2 - 2m_{H^\pm}^2 \right) \\[2mm]
\lambda_5 &=& \frac{1}{v^2} \left( \frac{m_{12}^2}{c_\beta s_\beta} - m_A^2 \right) \\[2mm]
\lambda_6 &=& \frac{1}{v_S^2} (m_{H_1}^2 R_{13}^2 + m_{H_2}^2 R_{23}^2 + m_{H_3}^2 R_{33}^2) \\[2mm]
\lambda_7 &=& \frac{1}{v v_S c_\beta} (m_{H_1}^2 R_{11}R_{13} + m_{H_2}^2 R_{21}R_{23} + m_{H_3}^2 R_{31}R_{33}) \\[2mm]
\lambda_8 &=& \frac{1}{v v_S s_\beta} (m_{H_1}^2 R_{12}R_{13} + m_{H_2}^2 R_{22}R_{23} + m_{H_3}^2 R_{32}R_{33})
\end{array}
\end{equation}

\section{Parameter Spaces: scans and constraints } 
Exploring the parameter space of each of the four types was carried out using $\mathtt{ScannerS} \ \mathtt{v.2}$ \cite{Muhlleitner:2020wwk}. We modify the package to enable Latin Hypercube Sampling (LHS) \cite{LHS1, LHS2}, which we use along with random sampling. This hybrid scanning technique ensures good coverage of the parameter spaces. Particularly, LHS divides each range into N sections, where N is the desired number of samples, and guarantees that combinations from different sections of each parameter are systematically sampled. With that in mind, we collect around 120K samples for the four types, around a third of which is LHS. The scanned parameters are shown in Table \ref{tab:parameter_ranges},
\begin{table}[H]
\caption{Parameter ranges in the four types, where mass and VEV variables are in GeV.}
\label{tab:parameter_ranges}
%\centering
\begin{tabularx}{\textwidth}{@{}lcc@{}}
\toprule
\textbf{Parameter} & \textbf{Min Value} & \textbf{Max Value} \\ \midrule
\( m_{H_a} \)       & 125.09            & 125.09            \\
\( m_{H_b} \)       & 30                & 1500              \\
\( m_{H_c} \)       & 30                & 1500              \\
\( m_A \)           & 30                & 1500              \\
\( m_{H^\pm} \)     & 580               & 1500              \\
\( \tan\beta \)     & 0.8               & 20                \\
\( c^2_{H_aVV} \)   & 0.9               & 1                 \\
\( c^2_{H_att} \)   & 0.8               & 1.2               \\
\(\text{sign}(R_{a3})\) & -1            & 1                 \\
\( R_{b3} \)        & -1                & 1                 \\
\( m_{12}^2 \)      & \( 10^{-3} \)     & \( 5 \times 10^5 \) \\
\( v_s \)           & 1                 & 3000              \\ \bottomrule
\end{tabularx}
\end{table}

Within $\mathtt{ScannerS}$, each point in the parameter spaces is subjected to a series of validity checks and required to pass the following constraints\footnote{$\mathtt{ScannerS}$ interfaces with obsolete versions: HiggsBounds 5 and HiggsSignals 2, hence we do not utilize them.}:
\begin{itemize}
    \item Boundedness: To ensure that the scalar potential is bounded from below as the fields approach infinity, the following conditions need to be met \cite{Muhlleitner:2016mzt}:
\begin{equation}
\begin{array}{c@{\quad}c@{\quad}c}
\begin{aligned}
& \lambda_1 > 0, \ \lambda_2 > 0, \ \lambda_6 > 0, \\
& \sqrt{\lambda_1 \lambda_6} + \lambda_7 > 0, \\
& \sqrt{\lambda_2 \lambda_6} + \lambda_8 > 0, \\
& \sqrt{\lambda_1 \lambda_2} + \lambda_3 + D > 0, \\
& \lambda_7 + \sqrt{\frac{\lambda_1}{\lambda_2}} \lambda_8 \geq 0,
\end{aligned}
&
\text{or} \ 
&
\begin{aligned}
& \lambda_1 > 0, \ \lambda_2  > 0, \ \lambda_6 > 0, \\
& \sqrt{\lambda_1 \lambda_6} > -\lambda_7 \;\geq\; \sqrt{\frac{\lambda_1}{\lambda_2}}\,\lambda_8, \\
& \sqrt{\lambda_2 \lambda_6} \;\geq\; \lambda_8 \;>\; -\sqrt{\lambda_2 \lambda_6}, \\
& \sqrt{\bigl(\lambda_7^2 - \lambda_1 \lambda_6\bigr)\bigl(\lambda_8^2 - \lambda_2 \lambda_6\bigr)}
   > \lambda_7 \lambda_8 - (D + \lambda_3)\,\lambda_6,
\end{aligned}
\end{array}
\end{equation}
where \( D = \lambda_4 - \lambda_5 \) if \(\lambda_4 > \lambda_5 \) and zero otherwise.
    \item Perturbative unitarity: Ensure that the largest eigenvalue corresponding to $2\times 2$ scattering matrices is below the upper limit $8\pi$.
\begin{equation}
\begin{aligned}
&|\lambda_3 - \lambda_4| < 8\pi,\\
&|\lambda_3 + 2 \lambda_4 \pm 3 \lambda_5| < 8\pi,\\
&\left| \frac{1}{2} \bigl( \lambda_1 + \lambda_2 + \sqrt{(\lambda_1-\lambda_2)^2 + 4 \lambda_4^2} \bigr) \right| < 8\pi,\\
&\left| \frac{1}{2} \bigl( \lambda_1 + \lambda_2 + \sqrt{(\lambda_1-\lambda_2)^2 + 4 \lambda_5^2} \bigr) \right| < 8\pi,\\
&|\lambda_7| < 8\pi,\quad |\lambda_8| < 8\pi,\\
&\frac{1}{2} \bigl|a_{1,2,3}\bigr| < 8\pi.
\end{aligned}
\end{equation}
where $a_{1,2,3}$ are roots of the cubic equation given in \cite{Muhlleitner:2016mzt} (Eq. 3.37).
    \item Vacuum stability (by $\mathtt{EVADE}$ \cite{EVADE}): To ensure that the EW vacuum is stable or at least metastable and in that case is long-lived. 
    \item B Physics: Stringent constraints arise from the following processes (see Table 2 and Figure 9 in \cite{Haller:2018nnx}):
\begin{equation}
  \mathcal{B}(B \to X_s \gamma) \;=\; 
  \bigl(3.32 \pm 0.15_{\mathrm{stat}+\mathrm{syst}}\bigr) \times 10^{-4}
  \;\pm\; 7\%\bigl(\mathrm{theo}\bigr).
\end{equation}
\begin{align}
  \mathcal{B}(B_s \to \mu^+\mu^-)_{\text{LHCb}} 
  &= (3.0^{+0.6}_{-0.5}) \times 10^{-9},\\[6pt]
  \mathcal{B}(B_d \to \mu^+\mu^-)_{\text{LHCb}} 
  &= (1.5^{+1.2}_{-1.0}) \times 10^{-10},
\end{align}
    \item Electroweak precision measurements: Restricting the oblique parameters S, T, and U, where for the simplified scenario \(U=0\) we have \cite{Haller:2018nnx}:
\begin{equation}
      S\big|_{U=0} = 0.04 \pm 0.08, 
  \quad
  T\big|_{U=0} = 0.08 \pm 0.07,
\end{equation}

with a correlation coefficient of \( +0.92 \).
\end{itemize}

Finally, and after imposing the previous constraints on the generated samples, we interface each type with $\mathtt{HiggsTools}$ (HT) \cite{Bahl:2022igd} via python \footnote{We provide the interfacing and analysis codes upon request through our Github page \cite{N2HDMTools}}, to confront the model with the latest Higgs data via HT subpackages: $\mathtt{HiggsSignals}$ $\mathtt{v.3}$ (HS) with the HS repository $\mathtt{v.1.1}$ and $\mathtt{HiggsBounds}$ $\mathtt{v.6}$ (HB) with the HB repository $\mathtt{v.1.6}$.

%%%%%%%%%%%%%%%%%%%%%%%%%%%%%%%%%%%%%%%%%%
\section{Results and Discussion} 
In this section we present the results of the best-fit analysis, as well as the analysis of the most relevant processes for constraining the non-SM Higgs bosons. 

\subsection{SM-like Higgs Signals}
One of the major tasks of the CMS and ATLAS detectors at the LHC is to precisely measure the couplings of the SM-like Higgs to fermions and gauge bosons, which is done by measuring the production and decay channels. This can quantify any deviations from the SM, and sets stringent limits on BSM models with a scalar boson that resembles the SM-like Higgs boson to some extent. Indeed, having the same mass is not sufficient to claim SM-like Higgs bosons in a given BSM extension. The predictions of branching ratios and production cross sections have to be within observed measurements. A practical way to set such limits is through the $\mu$-framework. For a specific production channel $i$ followed by a specific decay channel $f$:
\begin{equation}
   \mu_{if} = \frac{(\sigma_i \times \mathcal{B}_f)^{\text{obs}}}{(\sigma_i \times \mathcal{B}_f)^{\text{SM}}}. 
\end{equation}  

 CMS and ATALS provide data on each measured $\mu_i$ and $\mu_f$, as well as a combined $\mu$ for all measurements. The latest combined $\mu$ presented by the Particle Data Group (PDG) is \cite{ParticleDataGroup:2024cfk},
\begin{equation}
    \mu = 1.03 \pm 0.04.
\end{equation}

$\mathtt{HS}$ computes $\chi^2$ from the signal rates of a given model, normalized by the SM, as (Eq. 6 in \cite{Bechtle:2020uwn},
\begin{equation}
    \chi_\mu^2 = (\hat{\vec{\mu}} - \vec{\mu})^T \mathbf{C}^{-1}_{\vec{\mu}} (\hat{\vec{\mu}} - \vec{\mu}),
\end{equation}
where the $\vec{\mu}$ vectors contain individual signal strengths as predicted by the models, while $\hat{\vec{\mu}}$ represents the corresponding measurement, and $\mathbf{C}^{-1}_{\vec{\mu}}$ is a covariance matrix encoding uncertainties.
$\mathtt{HS}$ also computes $\chi^2_m$ for the mass of SM-like Higgs. Given that we set the mass of the lightest CP-even Higss to the observed SM-like Higgs, only $\chi^2_\mu$ contributes, hence we drop the subscript in the subsequent analysis. As a reference value, the SM with $m_h = 125.09$ GeV gives $\chi^2 = 152.54$, as calculated by $\mathtt{HS}$ with 159 observables. For a given point ($p$) in N2HDM, we define:
\begin{equation}
    \Delta \chi^{2} = \chi^{2}_p - {\chi^{2}_{\text{min}}},
\end{equation}  
where $\chi^2_{\text{min}}$ is the minimum value in the parameter space, representing the best-fit point.

To facilitate comparison with 2HDM, we present the results by shifting $\alpha_1$ by $-\frac{\pi}{2}$ so that the shifted angle is equivalent to the 2HDM convention. In this case, 
\begin{equation}
    c_{H_1VV} = \cos{(\beta - \alpha_1)}\cos{(\alpha_2)} \xrightarrow{\alpha_1 - \frac{\pi}{2}} \tilde{c} \equiv \sin{(\alpha_1 - \beta)}\cos{(\alpha_2)}.
\end{equation}

We note that, in the allowed data for all types, $\cos{\alpha_2} \sim \mathcal{O}(1)$. However, we emphasize that we are not necessarily in the 2HDM limit of the N2HDM since $\alpha_3$ is not always close to zero in the valid parameter space.

\begin{figure}[H]
    \centering    
    \begin{subfigure}[b]{0.47\textwidth}
        \centering
        \includegraphics[width=\textwidth]{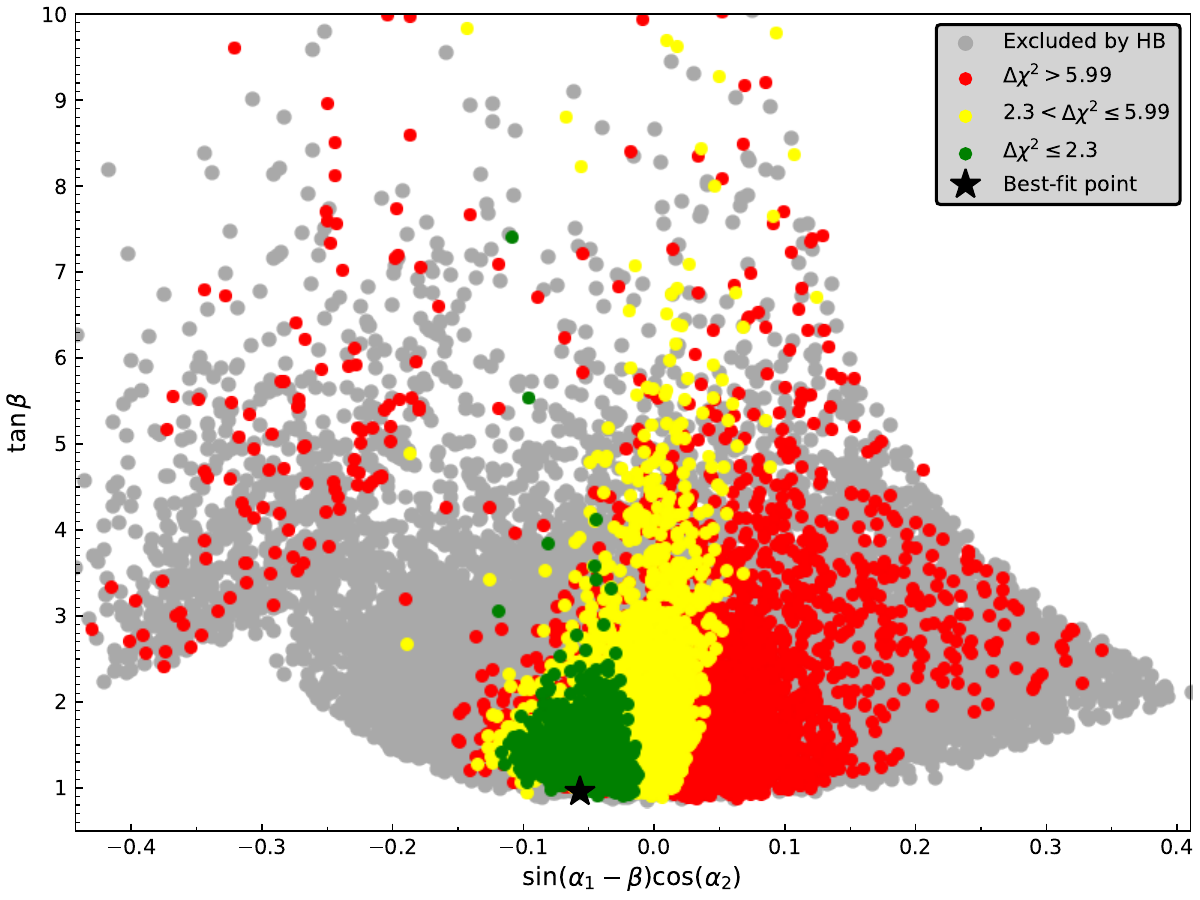}
        \caption{Type 1}
        \label{fig:type1}
    \end{subfigure}
    \hfill
    \begin{subfigure}[b]{0.47\textwidth}
        \centering
        \includegraphics[width=\textwidth]{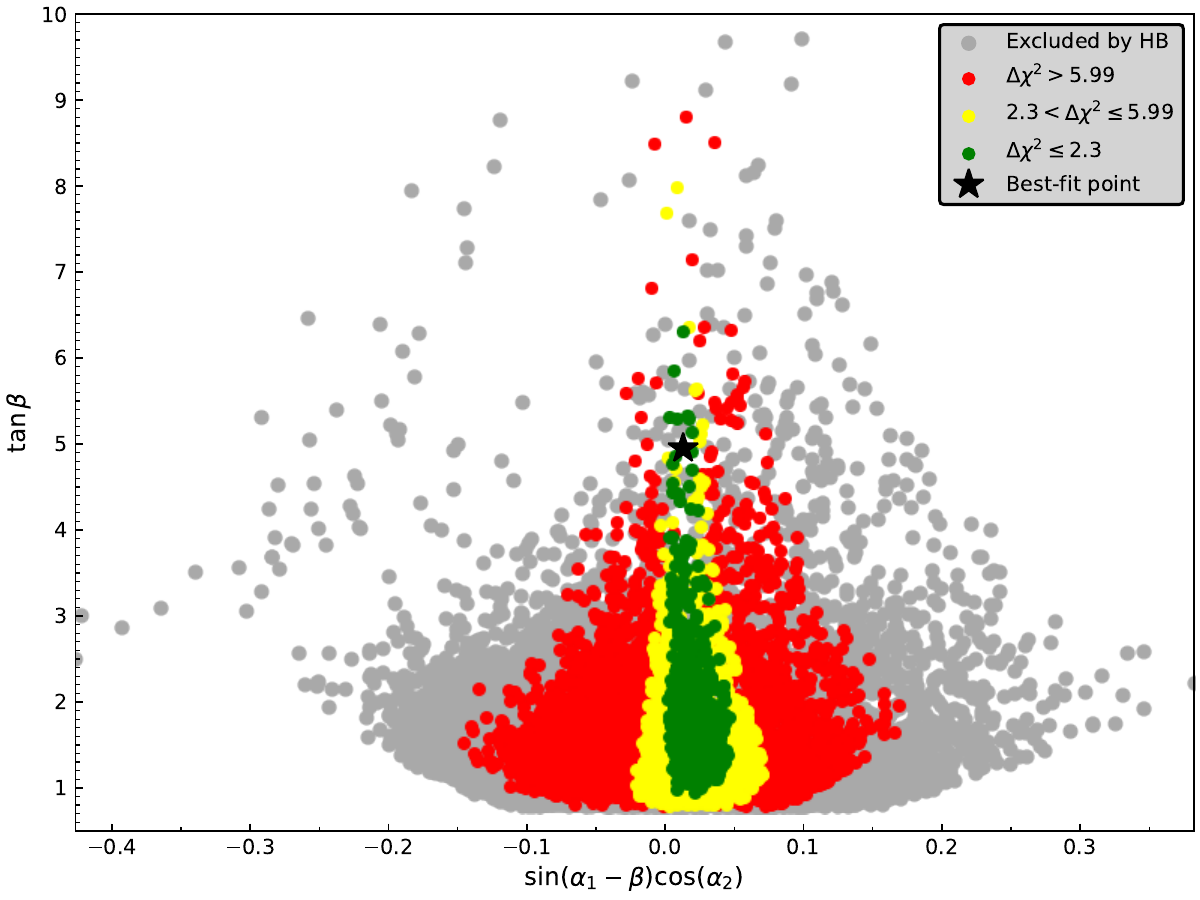}
        \caption{Type 2}
        \label{fig:type2}
    \end{subfigure}
    \vspace{0.5cm} 
    \begin{subfigure}[b]{0.47\textwidth}
        \centering
        \includegraphics[width=\textwidth]{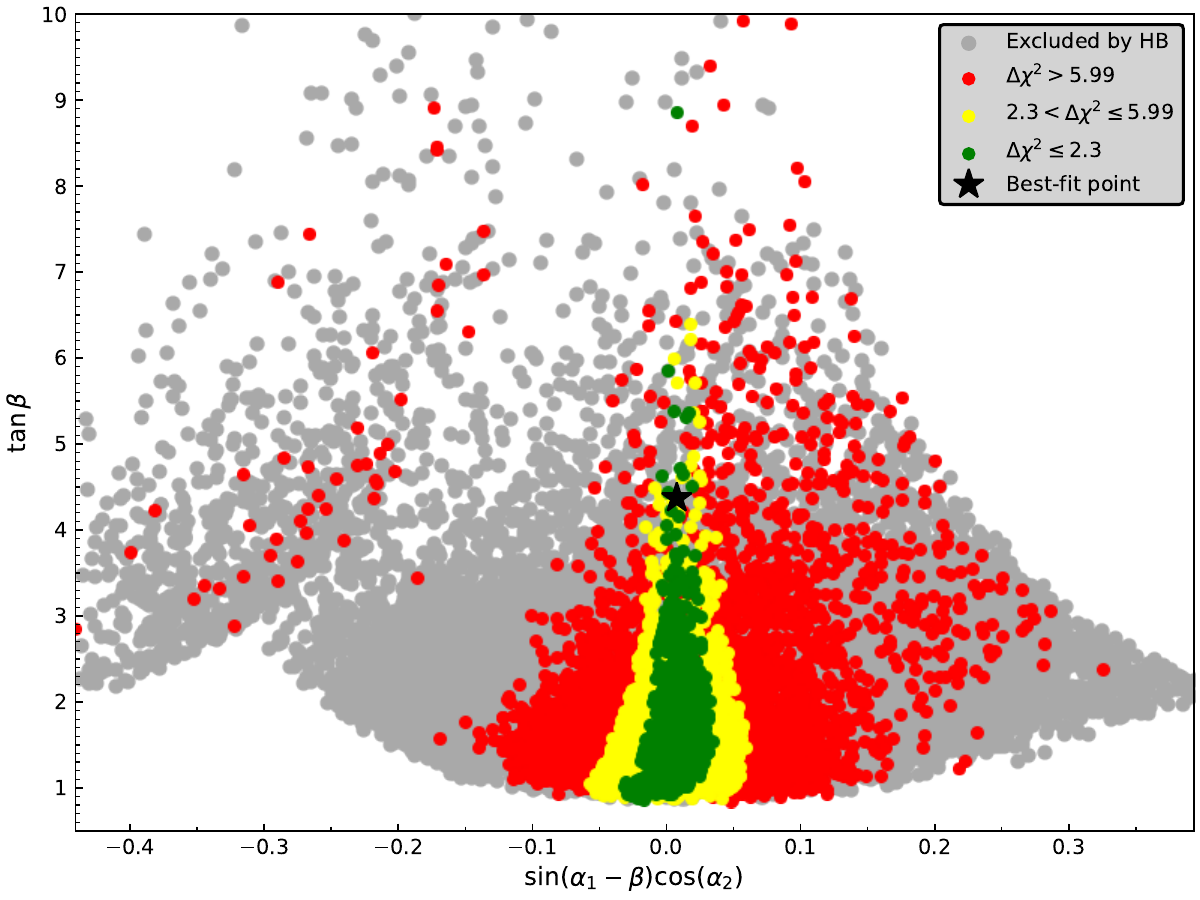}
        \caption{Type X}
        \label{fig:type3}
    \end{subfigure}
    \hfill
    \begin{subfigure}[b]{0.47\textwidth}
        \centering
        \includegraphics[width=\textwidth]{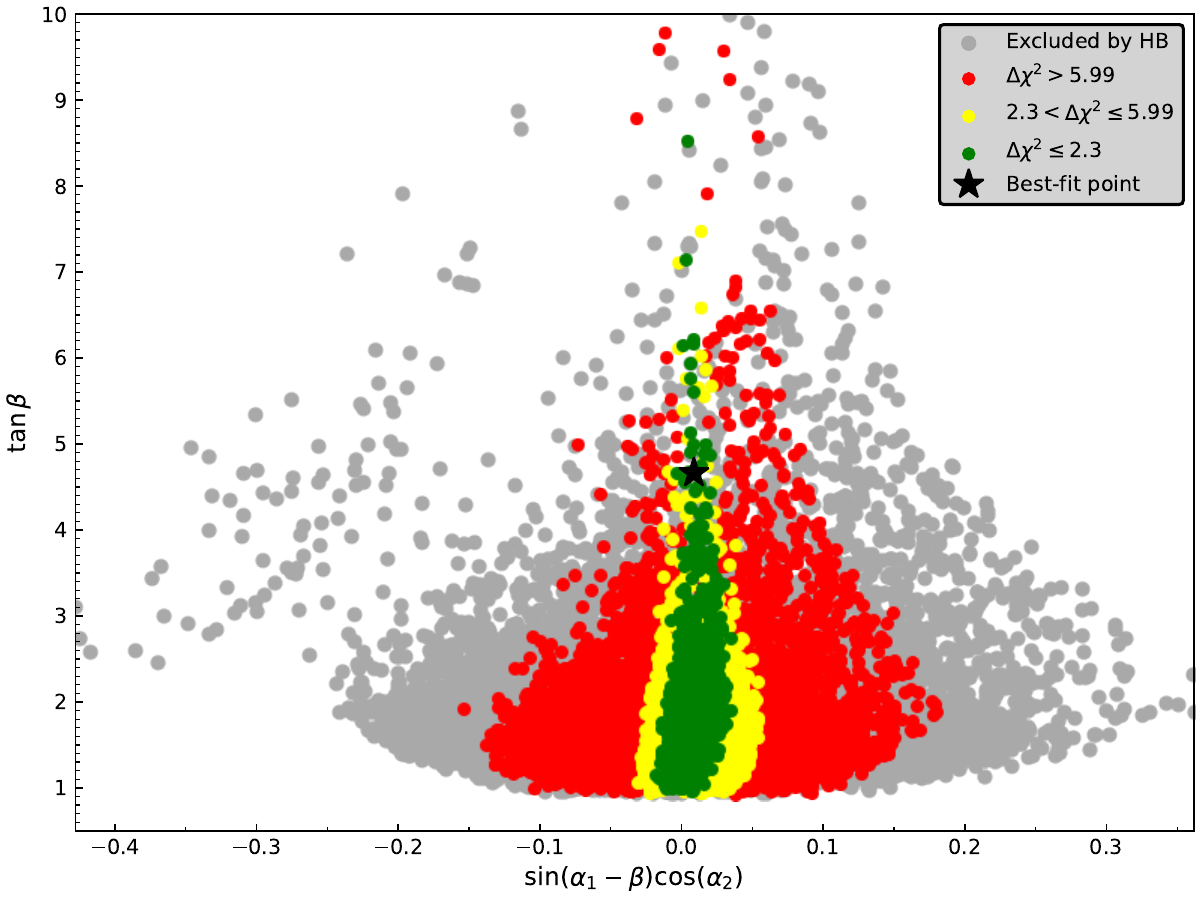}
        \caption{Type Y}
        \label{fig:type4}
    \end{subfigure}
    \caption{$\chi^2$ analysis in the $\tilde{c}(H_1 VV)$-$\tan\beta$ planes.}
    \label{fig:N2HDM_results_all}
\end{figure}

Figure~\ref{fig:N2HDM_results_all} shows $\mathtt{HS}$ results and the $\chi^2$ analysis for the four types of N2HDM. Points that are colored red are ruled out by $\mathtt{HS}$ since $\Delta \chi^2 > 5.99$. The yellow points represent the $95\%$ C.L. for which $\Delta \chi^2 \leq 5.99$ (see Table 4 in \cite{Bechtle:2020uwn}), while the green points are for cases where $\Delta \chi^2 \leq 2.3$, which fall in the $68\%$ C.L. The black stars represent the best-fit points where $\Delta \chi^2$ is minimum. For completeness, the gray points represent cases that are ruled out by $\mathtt{HB}$. 

For T1, we observe in the top-left panel of Figure~\ref{fig:N2HDM_results_all}, that the best-fit point corresponds to a small value of $\tan \beta = 0.96$, and that it is slightly shifted to the negative side $\tilde{c} = -0.057$, which coincides with the results of the 2HDM presented in \cite{ATLAS:2021vrm} and the case study in \cite{Bahl:2022igd}. Most of the green points reside between $ -0.1 < \tilde{c} < 0$ with $ 0 < \tan{\beta} < 3$, while the yellow points expand to the positive side where $\tilde{c} > 0.05$, especially as $\tan{
\beta}$ becomes larger than 2. As $\tan{\beta}$ becomes smaller, the positive side of $\tilde{c}$ is disallowed by \texttt{HS} as indicates by the red points.

Next, in T2, which is shown in the top-right panel, the best-fit point is located at $\tilde{c} = 0.013$, and corresponds to $\tan \beta = 4.95$, which is larger than T1. Points that fall within the $68\%$ C.L. region are located on the positive side where $\tilde{c} \sim 0.05$ for $\tan{\beta} < 2$, and approach zero as $\tan{\beta}$ becomes larger. On the other hand, the yellow region extends from $0.02 < \tilde{c} < 0.07$ for small values of $\tan{\beta}$, and converges near $\tilde{c} \sim 0$ as $\tan{\beta}$ increases.

As for TX, in the bottom-left panel, we can see that the green points shifted slightly to negative for smaller values of $\tan \beta$ where $\tilde{c} \sim -0.04$, and as $\tan{\beta}$ increases, this region becomes more symmetric around $\tilde{c} = 0$. The yellow points, on the other hand, extend between $-0.06 < \tilde{c} < 0.6$ and $\tan{\beta} < 2$, and shirk as it increases. The best-fit point lies at $\tan \beta = 4.37$, for which $\tilde{c}= 0.008$. 

In TY, as shown in the bottom-right panel, the allowed region is centered at zero. Specifically, the green points span $-0.01 < \tilde{c} < 0.04$, for $0.8 < \tan{\beta} < 3$, while the yellow points extend slightly further and up to $\tilde{c} \sim 0.05$. Both regions become closer to $\tilde{c} = 0$ as $\tan{\beta}$ increases. The best-fit point has $\tan \beta = 4.66$ and $\tilde{c} = 0.009$.

To better understand the specific measurements affecting regions surrounding the best-fit points, we compute, using $\mathtt{HS}$, individual $\chi^2$ for two points at $\tan\beta_{\chi^2_{\text{min}}} \pm \delta$, where $\delta \leq \tan \beta_{\chi^2_{\text{min}}} \times 2\%$, and $\tilde{c} = \tilde{c}_{\chi^2_{\text{min}}} \pm \gamma$, where $\gamma \leq 0.05$, depending on the availability of the data. Then we compute $\Delta \chi^2$ between the two selected points to find measurements that lead to significant deviations from the value of $\chi^2_{\text{min}}$. We find that these regions are mostly affected by searches for:
\begin{itemize}
    \item $pp \rightarrow h \rightarrow VV \rightarrow 4l$ \cite{ATLAS:2020rej, CMS:2021ugl, CMS:2020dvg, ATLAS:2018xbv},
    \item $pp \rightarrow ht\bar{t} \rightarrow (h\rightarrow b \bar{b}) (t/\bar{t} \rightarrow \textit{semileptonic})$ \cite{ATLAS:2021qou},
    \item $pp \rightarrow Vh \rightarrow (h\rightarrow b \bar{b}) (V \rightarrow ll/l\nu/\nu\nu)$ \cite{CMS:2018nsn, ATLAS:2020fcp},
    \item $pp \rightarrow h \rightarrow \tau^+ \tau^-$ \cite{CMS:2021sdq, CMS:2022kdi},
    \item $pp \rightarrow h \rightarrow \gamma \gamma$ \cite{ATLAS:2022tnm, CMS:2021kom}
\end{itemize}

Figure~\ref{fig:N2HDM_results_all} also indicates that deviations from the SM are allowed in the four types; however, this is more pronounced in T1, for which the allowed region does not converge sharply to the alignment limit with the increase of $\tan\beta$ as is the case in the other three types. This distinct feature is common in 2HDMs as indicated previously. However, it should be mentioned that the properties of SM-like Higgs in N2HDM can differ from 2HDM due to the presence of an additional singlet component $|R_{13}|^2$. Particularly, in terms of the model parameters, we observe that, in T1 and TX, the singlet component of SM-like Higgs is restricted to values below $10\%$, while it is below $15\%$ in T2, and below $18\%$ in TY. Furthermore, one can set upper and lower limits on the angles $\alpha_1$ and $\alpha_2$ based on the allowed results, as shown in Table ~\ref{tabalpha}, while $\alpha_3$ is found to be allowed to take the full range specified in the previous Section.

\begin{table}[H]
%\centering
\caption{Allowed ranges of the mixing angles $\alpha_1$ and $\alpha_2$.}
\label{tabalpha}
\begin{tabularx}{\textwidth}{|c|c|c|}
\hline
\text{} & $\alpha_1$ (min, max) & $\alpha_2$ (min, max) \\
\hline
Type 1 & $-1.556$, $1.563$ & $-0.3139$, $0.2787$ \\
Type 2 & $0.6649$, $1.502$ & $-0.4026$, $0.3853$ \\
Type X & $0.6908$, $1.466$ & $-0.3115$, $0.2923$ \\
Type Y & $0.7366$, $1.469$ & $-0.4235$, $0.4339$ \\
\hline
\end{tabularx}
\end{table}

\subsection{Bounds on the Additional CP-Even Higgs Bosons}
ATLAS and CMS systematically search for non-SM Higgs bosons decaying into lighter bosons or fermions in several final states. This case is relevant for this work, since we are considering the ordering $m_{H_1}^{\text{SM}} < m_{H_2} < m_{H_3}$. However, experimental results usually assume NWA, hence they can set limits on the additional Higgs bosons in the N2HDM provided that $\frac{\Gamma_{H}}{m_{H}} \ll 1$, where $\Gamma$ is the total width of the additional Higgs. Nevertheless, certain searches presented results valid beyond NWA, and these are taken into account, as will be discussed later. For each additional Higgs, $\mathtt{HT}$'s subpackage $\mathtt{HiggsPredictions}$ ($\mathtt{HP}$) computes the production cross-sections in the effective coupling approximation, including QCD corrections. We take all main production channels into account, especially that Vector Boson Fusion (VBF) and production in association with a vector boson (HV) can be larger than Gluon Fusion (ggH) in regions where the coupling of the additional Higgs to up-type quarks is very small compared to its coupling to gauge bosons. Similarly, production in association with bottom and anti-bottom quarks (bbH) can be larger than ggH in certain regions. All these effects have been taken into account in our analysis, within the precision provided by $\mathtt{HT}$ and its subpackages. 

Furthermore, for each instance of input parameters, $\mathtt{HB}$ determines the most sensitive measurement. This is done by computing the model's prediction for each relative $\sigma \times \mathcal{B}_i$ divided by the corresponding expected limits. The limit that maximizes this ratio is considered the most sensitive. Next, it computes the observed ratio (model's prediction for a specific decay channel divided by the corresponding observed limits). If the ratio is larger than 1, then it determines that the point is excluded at the $95\%$ C.L. \cite{Bechtle:2008jh, Bechtle:2020pkv}. 

For each type, we discuss the most sensitive channels based on $\mathtt{HB}$, paying special attention to heavy Higgs resonances decaying into a pair of bosons or fermions. All points already pass constraints from $\mathtt{HS}$ and the set of constraints described in the previous Section. Since the four types are generally affected by a common set of measurements, we present the results with respect to each class of relevant measurements. Moreover, in the presented figures, we use the HEP Inspire biographical code. 

The most stringent bounds on the additional CP-even Higgs bosons come from the classes of searches listed below, where we denote $H_1^{\text{SM}} \equiv h$ and $H$ is an additional non-SM Higgs boson.

\subsubsection{Class: $p p \rightarrow H \rightarrow h h$ }
\begin{figure}[H]
    %\centering
    \includegraphics[width=0.8\linewidth]{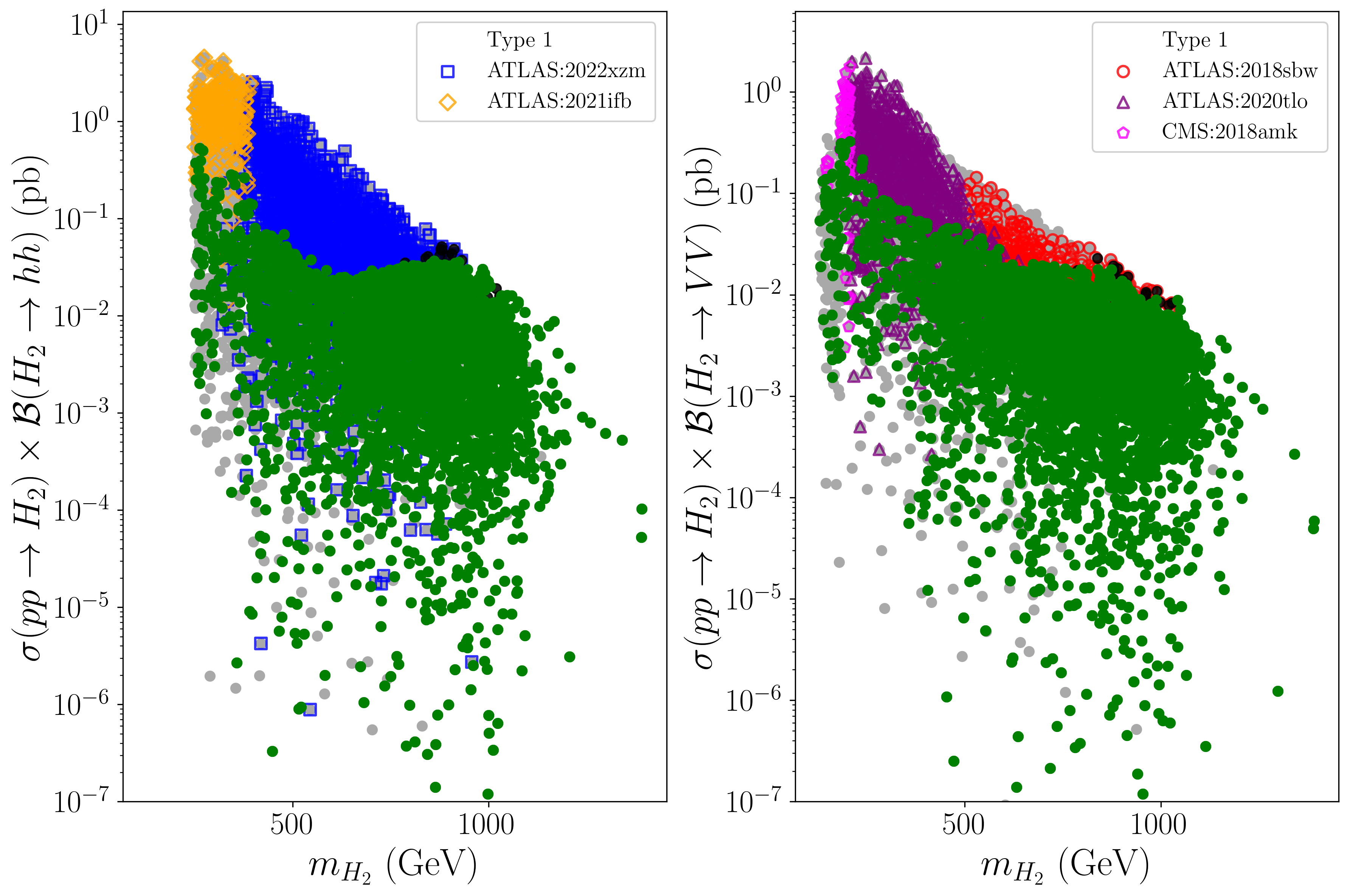}
    \caption{Production cross section times branching ratio to bosons for $H_2$. }
    \label{fig:H2_t1}
\end{figure}

\begin{figure}[H]
    %\centering
    \includegraphics[width=0.8\linewidth]{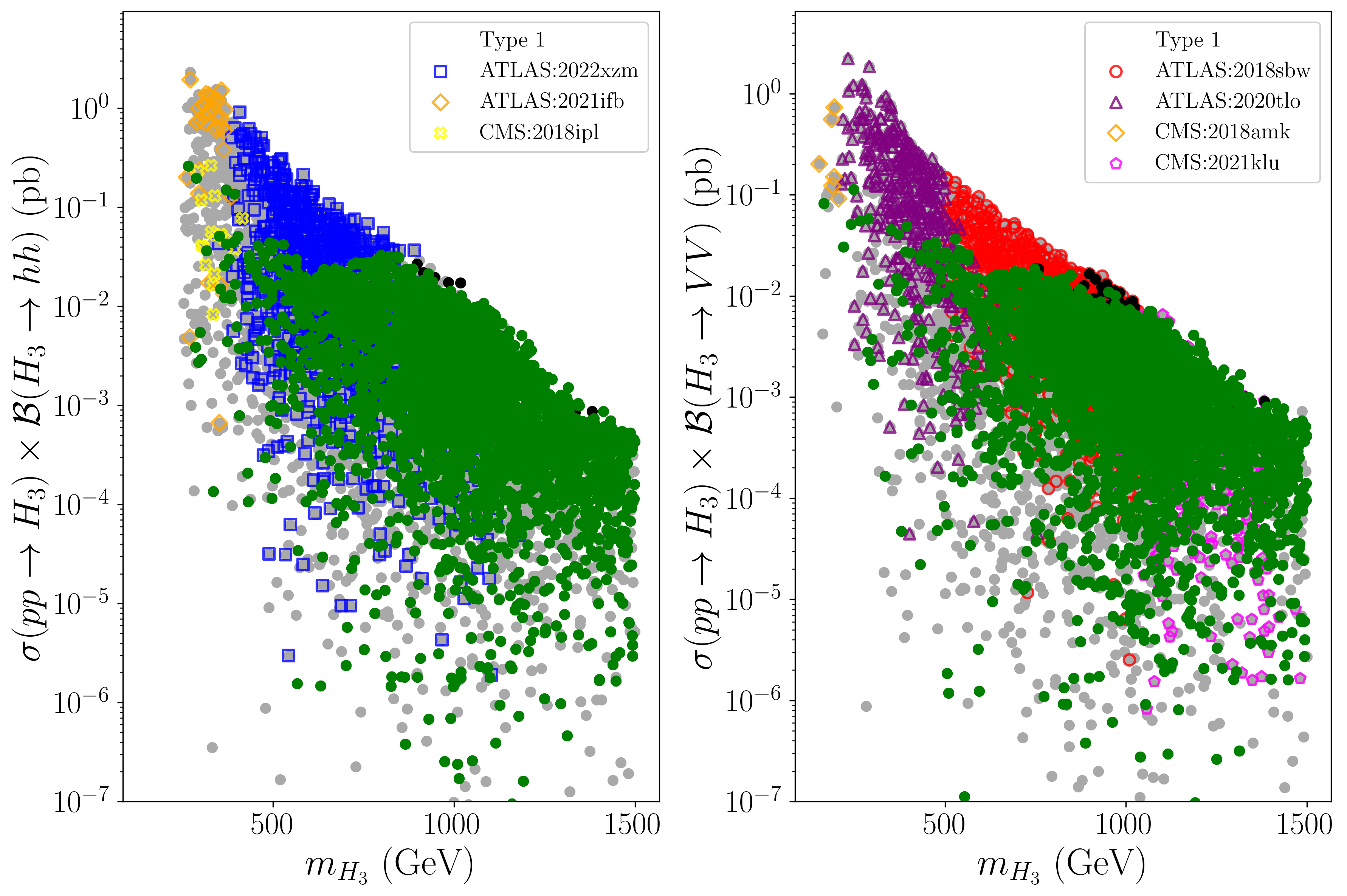}
    \caption{Production cross section times branching ratio to bosons for $H_3$.}
    \label{fig:H3_t1}
\end{figure}

\begin{figure}[H]
    %\centering
    \includegraphics[width=0.8\linewidth]{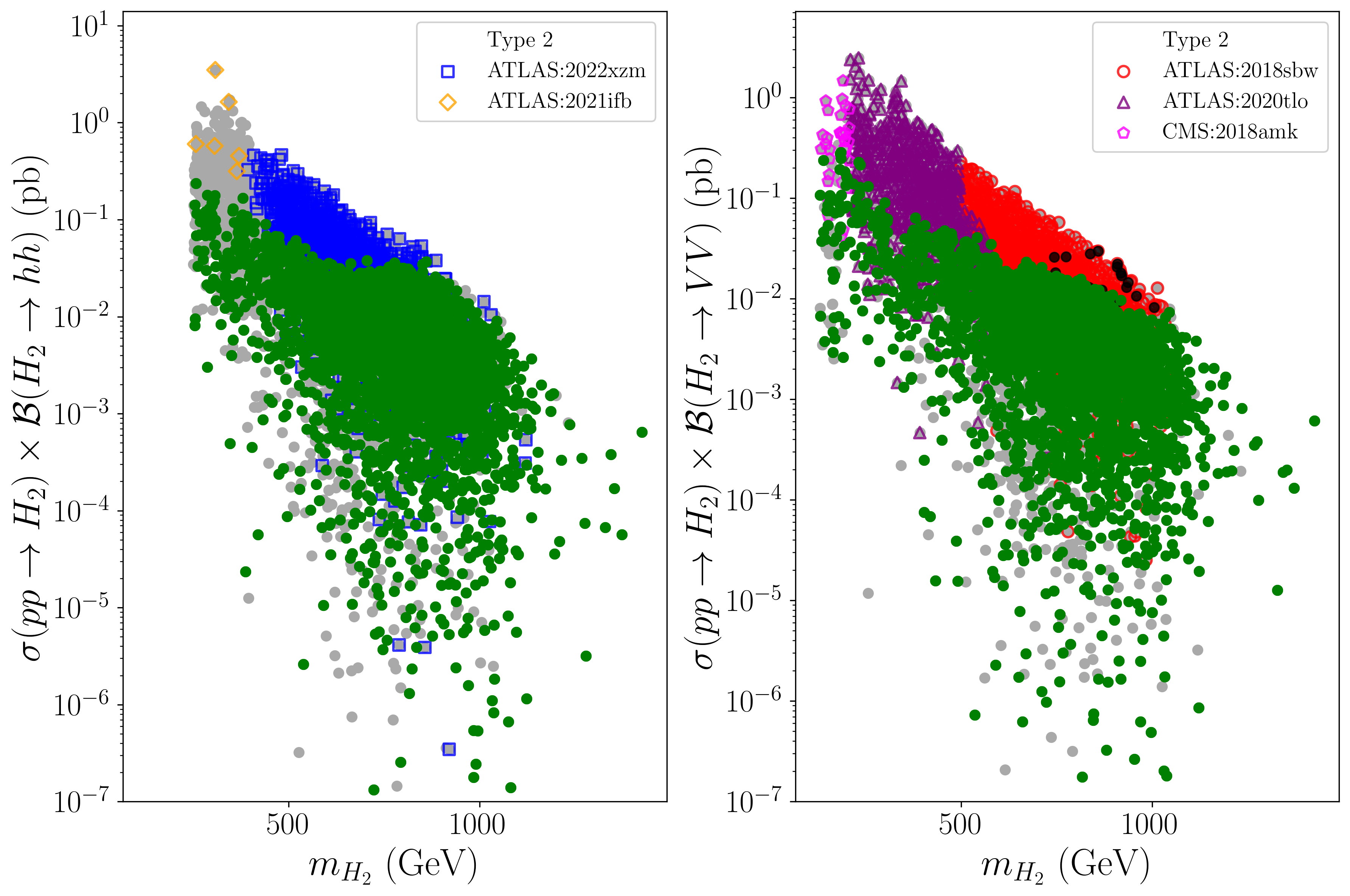}
    \caption{Production cross section times branching ratio to bosons for $H_2$.}
    \label{fig:H2_t2}
\end{figure}

\begin{figure}[H]
    %\centering
    \includegraphics[width=0.8\linewidth]{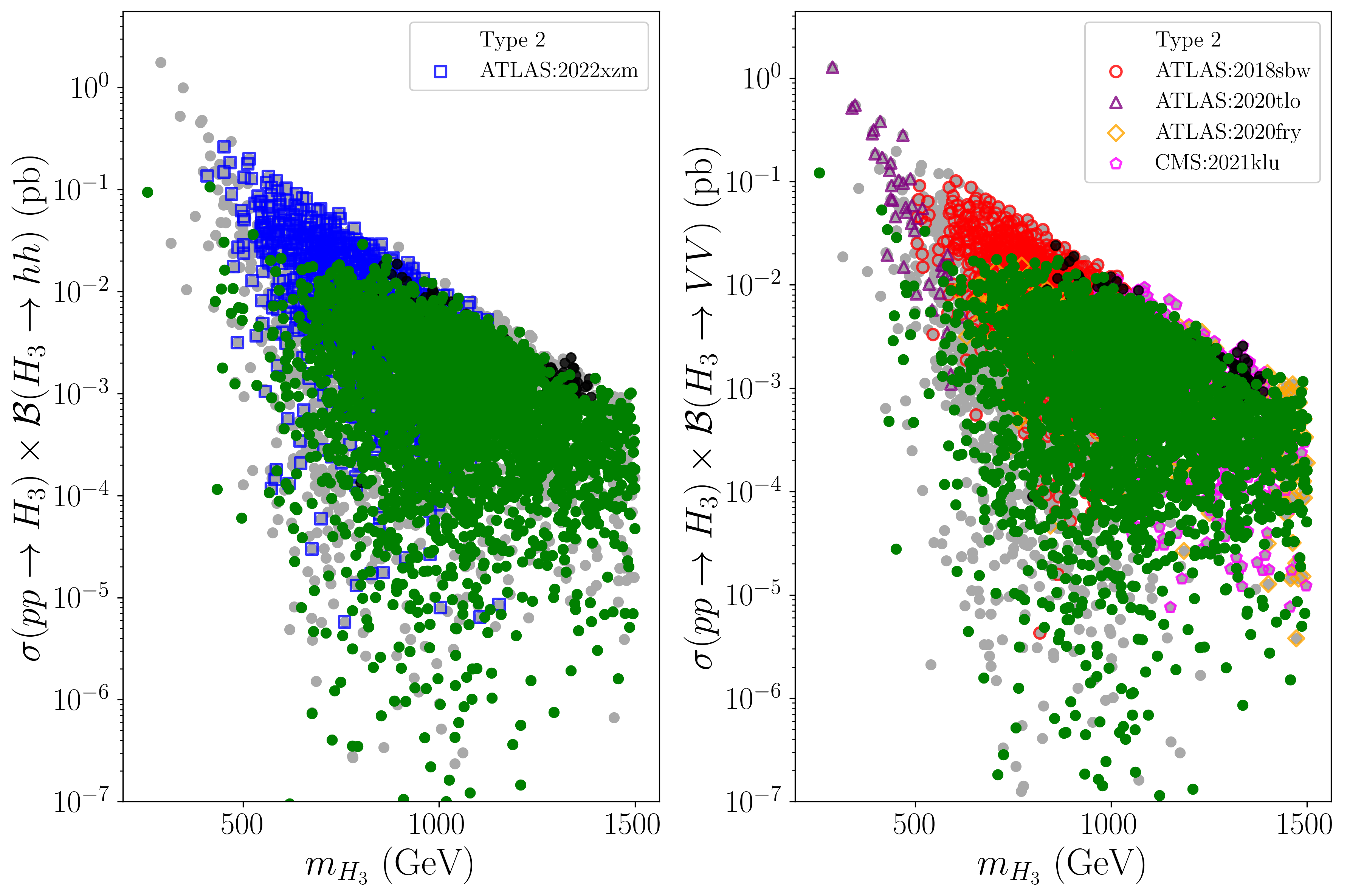}
    \caption{Production cross section times branching ratio to bosons for $H_3$.}
    \label{fig:H3_t2}
\end{figure}

An important class of LHC searches is that of an additional Higgs decaying into two SM-like Higgs bosons. In the analyzed parameter spaces of the four types, the most sensitive ones are:

\begin{itemize}
    \item $p p \rightarrow H \rightarrow h h \rightarrow \tau^- \tau^+ b \bar{b}$ \cite{ATLAS:2022xzm}
    \item $ p p \rightarrow H \rightarrow h h \rightarrow b\bar{b} \gamma \gamma$ \cite{ATLAS:2021ifb}
    \item $ p p \rightarrow H \rightarrow h h \rightarrow \textit{fermions/bosons}$ \cite{CMS:2018ipl}
\end{itemize}
where the results corresponding to this class appear in the left panels of Figures \ref{fig:H2_t1}-\ref{fig:H3_ty}. In these panels, green points are allowed, and grey points are ruled out by a different class (i.e. $pp \rightarrow H \rightarrow VV$). 

We observe that the limit from ATLAS (139 $fb^{-1}$) on narrow resonance production of a pair of SM Higgs bosons in the $\tau^- \tau^+ b \bar{b}$ final state ~\cite{ATLAS:2022xzm} ($\texttt{ATLAS:2022xzm}$), places strong constraints as it contributes the most to ruling out points in the parameter spaces. This constraint is relevant for a mass range of $H_2/H_3$ between 251 and 1600 GeV. It is very sensitive to $hh$ searches, given the relatively low background accompanied by a branching ratio of $\mathcal{B}(h h \rightarrow \tau^- \tau^+ b \bar{b} ) \sim 0.073$. The upper limits observed in $\sigma \times \mathcal{B}$ range from 0.9 pb to 0.021 pb. As can be seen in the left panels of Figures \ref{fig:H2_t1}-\ref{fig:H3_ty}, all types are affected by this measurement, and the corresponding ruled out points appear in blue. In particular, we observe that this search is deemed to be the most sensitive for values of mass above 400 GeV. Regions below that are affected by other searches that will be discussed shortly. Furthermore, we note that in T1, for $H_2$ with a mass around 900 GeV, Figure \ref{fig:H2_t1} (left panel) shows a few black points where $\texttt{HB}$ selects this search to be the most sensitive; however, these points reside beyond NWA.

Moreover, ATLAS (139 $fb^{-1}$) provided limits on new hypothetical heavy scalars in the mass range 251 GeV to 1000 GeV, which are set through searches in the final states to $b\bar{b}$ quarks and a pair of photons \cite{ATLAS:2021ifb} ($\texttt{ATLAS:2021ifb}$). The observed upper limits range from 0.64 pb to 0.044 pb. This search is responsible for ruling out parameter points (orange) with large $\sigma\times \mathcal{B}$ and $m_{H_2} < 400$ GeV, as seen in the left panels of Figures \ref{fig:H2_t1} (T1), \ref{fig:H2_t2} (T2) and \ref{fig:H2_tx} (TX).

Finally, for this class, limits on a new heavy scalar boson in the mass range between 270 and 3000 GeV are obtained from the results of CMS (35.9 fb$^{-1}$) \cite{CMS:2018ipl} ($\texttt{CMS:2018ipl}$) which searched for a new scalar decaying into two SM-like Higgs bosons, one of which subsequently decays into $b\bar{b}$, while the other one could decay to $b\bar{b}/\tau^+\tau^-/VV/\gamma \gamma$. The observed upper limits range from 0.68 pb to 0.002 pb. The effect of this measurement is visible in the left panel of Figure \ref{fig:H3_t1} for $H_3$ of T1, and appears in yellow, where the mass is below 400 GeV.


\subsubsection{Class: $ p p \rightarrow H \rightarrow VV$ }
\begin{figure}[H]
    %\centering
    \includegraphics[width=0.8\linewidth]{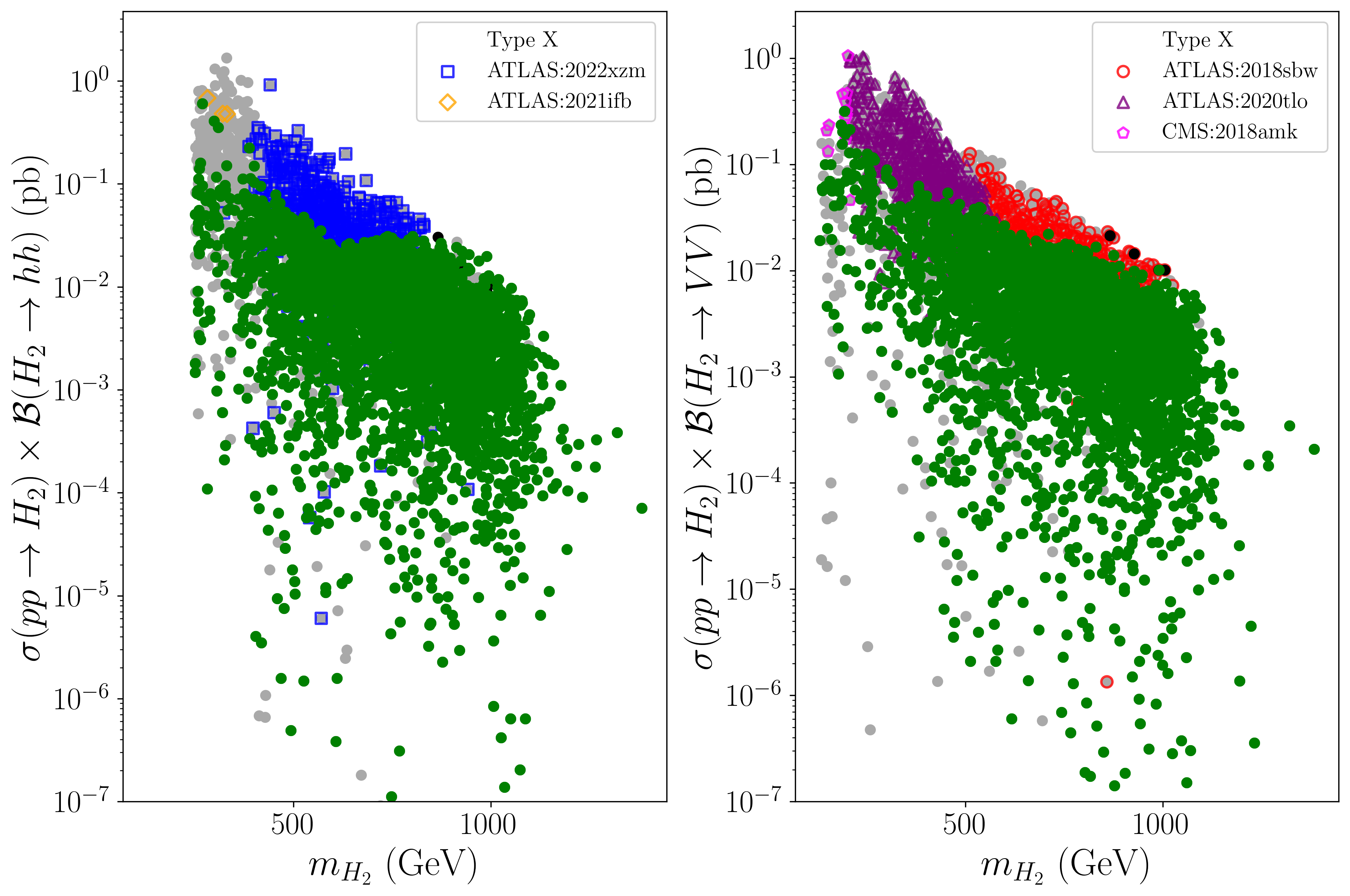}
    \caption{Production cross section times branching ratio to bosons for $H_2$.}
    \label{fig:H2_tx}
\end{figure}

\begin{figure}[H]
    %\centering
    \includegraphics[width=0.8\linewidth]{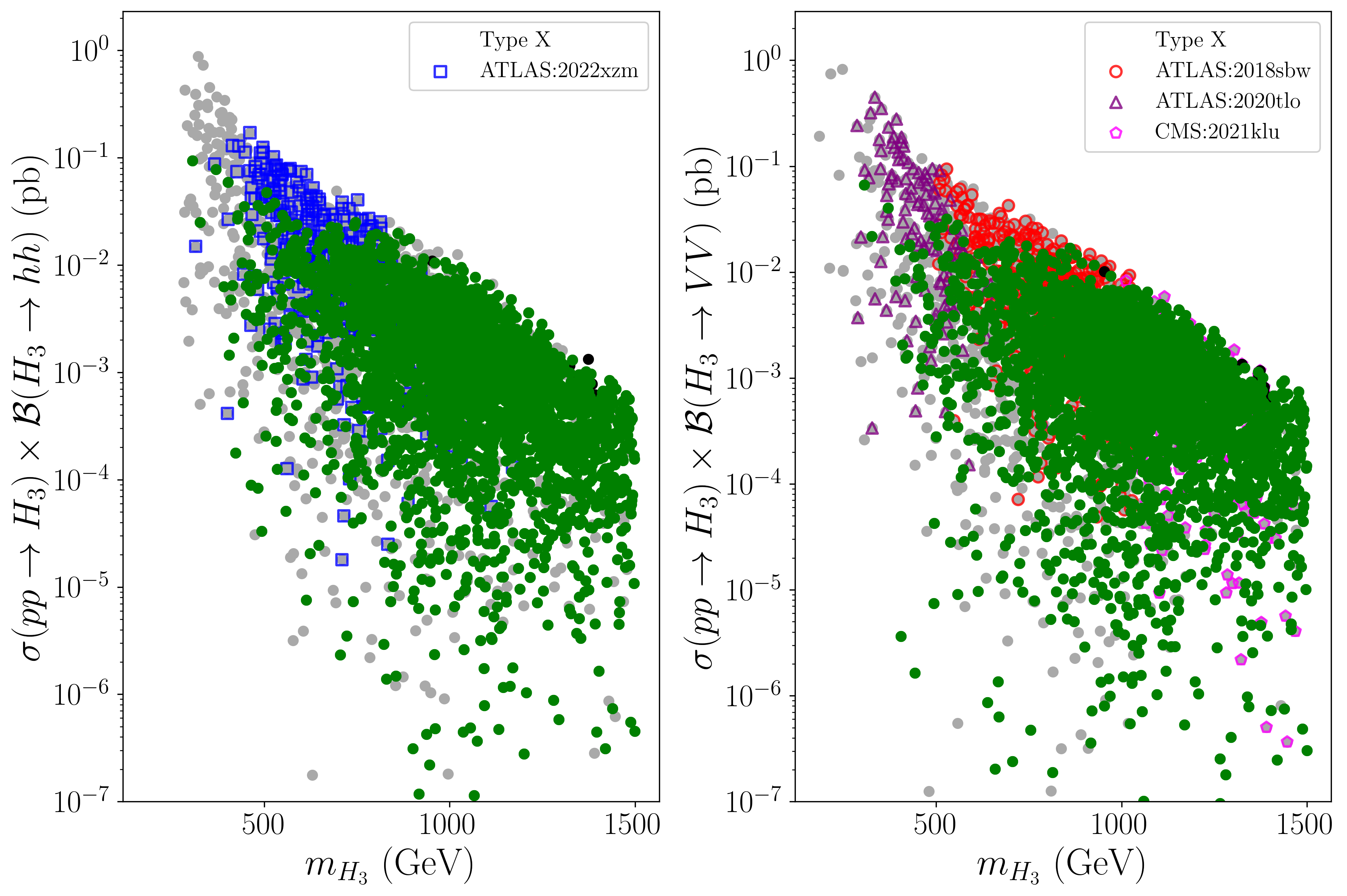}
    \caption{Production cross section times branching ratio to bosons for $H_3$.}
    \label{fig:H3_tx}
\end{figure}

\begin{figure}[H]
    %\centering
    \includegraphics[width=0.8\linewidth]{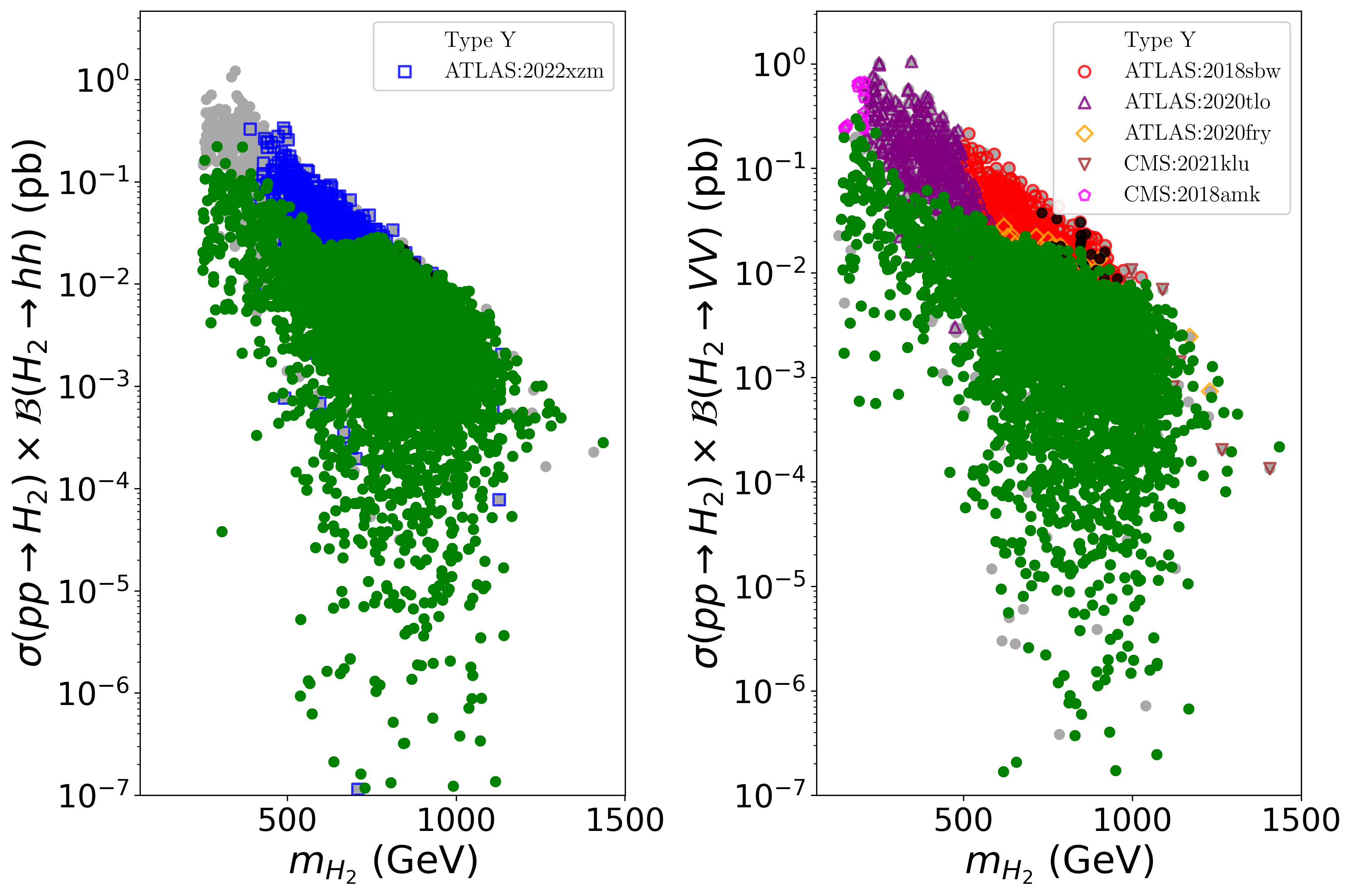}
    \caption{Production cross section times branching ratio to bosons for $H_2$.}
    \label{fig:H2_ty}
\end{figure}

\begin{figure}[H]
    %\centering
    \includegraphics[width=0.8\linewidth]{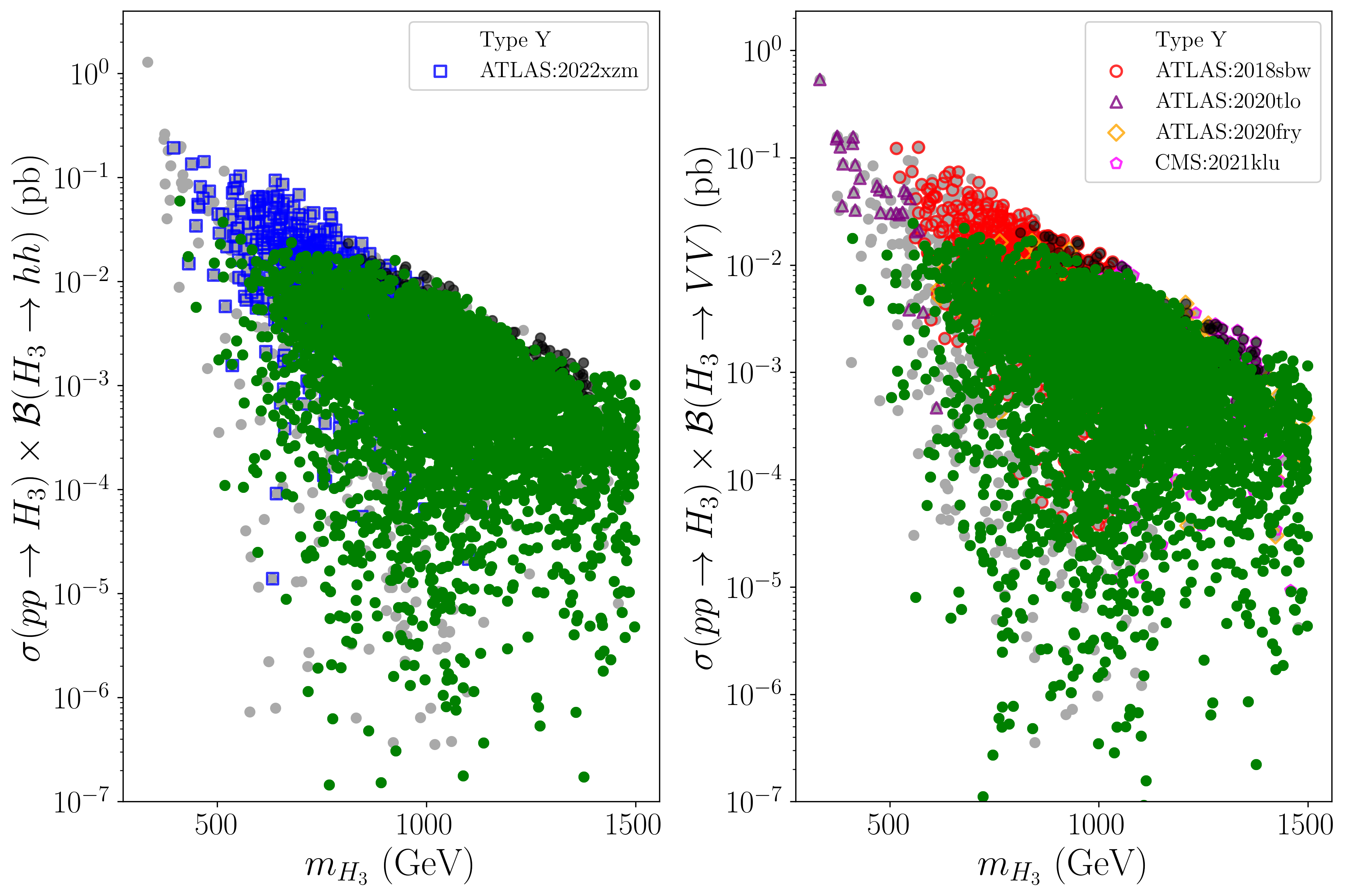}
    \caption{Production cross section times branching ratio to bosons for $H_3$.}
    \label{fig:H3_ty}
\end{figure}

A general search conducted at the LHC is that for a heavy resonance decaying into a pair of gauge bosons, which would subsequently decay into fermions. While some searches considered different fermionic final states, some searches focused on semileptonic or leptonic final states. The majority of constraints come from this class of measurements, and we find that the most sensitive ones are: 
\begin{itemize}
    \item $ p p \rightarrow H \rightarrow VV \rightarrow \textit{fermions}$ 
    \cite{ATLAS:2018sbw,ATLAS:2020tlo,CMS:2018amk}

    \item $pp \rightarrow H \rightarrow VV/Vh \rightarrow \textit{semileptonic}$
    \cite{CMS:2021klu,ATLAS:2020fry}
\end{itemize}
where the results corresponding to this class appear in the right panels of Figures \ref{fig:H2_t1}-\ref{fig:H3_ty}. In these panels, green points are allowed, and grey points are ruled out by a different class (i.e. $pp \rightarrow H \rightarrow hh$).

ATLAS (36 $fb^{-1}$) performed a general search for a heavy scalar resonance producing two bosons, which subsequently decay into fermions \cite{ATLAS:2018sbw} ($\texttt{ATLAS:2018sbw}$), this search is relevant for a mass range between 300 GeV and 3000 GeV. For ggH, the observed upper limits range from 0.38 (300 GeV) to 0.0013 (3 TeV), while for VBF, the observed upper limits range from 0.13 (500 GeV) to 0.0033 (3 TeV). The effects on the four types are shown in red, specifically in the right panels of each plot corresponding to $H \rightarrow VV$ in Figures \ref{fig:H2_t1}-\ref{fig:H3_ty} where $m_H \geq 500$ GeV. Some black points where $m_H > 700$ GeV, for which this measurement was deemed the most sensitive, do not satisfy the NWA and are shown in the right panels of Figures \ref{fig:H2_t1} and \ref{fig:H3_t1} (T1), \ref{fig:H2_t2} and \ref{fig:H3_t2} (T2), \ref{fig:H2_tx} (TX), \ref{fig:H2_ty} and \ref{fig:H3_ty} (TY).

Moreover, ATLAS (139 $fb^{-1}$) presented results concerning the production of a resonant non-SM Higgs ($H_2/H_3$) decaying into two gauge bosons, which subsequently decay into leptons \cite{ATLAS:2020tlo} ($\texttt{ATLAS:2020tlo}$). In models with two doublets, the upper limits for ggH and in the mass range between 200 GeV and 400 GeV is given in the NWA, where the range is from 0.11 pb to 0.047 pb. For larger mass values limits beyond the NWA are included up to $\frac{\Gamma}{m_H} = 0.15$. For VBF, the upper limits range from 0.031 pb (210 GeV) to 0.0017 pb (2 TeV). All points affected by this measurement are shown in purple, where we can see that the effects are severe in the mass regions below 600 GeV in the right panels of Figures \ref{fig:H2_t1}-\ref{fig:H3_ty}.

Moreover, the search by CMS \cite{CMS:2018amk} 
($\texttt{CMS:2018amk}$) at 35.9 $fb^{-1}$ on new resonances decaying into ZZ bosons, which subsequently decay into $4l$, $2l2q$, or $2l2\nu$, is valid for mass regions between 130 GeV and 3 TeV. It covers a wide ragne of width: $0 \leq \frac{\Gamma}{m_H} \leq 0.3$. For ggH, in the NWA, the upper limits range from 0.235 pb (130 GeV) to 0.0012 pb (3 TeV), while for $\frac{\Gamma}{m}$ = 0.3 they range from 0.104005 pb (130 GeV) to 0.006 pb (3 TeV).
For VBF, they range from 0.166 pb (130) to 0.0011 pb (3 TeV). As for larger width where $\frac{\Gamma}{m} = 0.3$, the range is between 0.0243 pb (130) to 0.0018 pb (3 TeV). We can see in the right panels of Figures \ref{fig:H2_t1}-\ref{fig:H3_ty} (pink points), that this measurement rules out some vertical points corresponding to a mass range below 200 GeV. 

Additionally, CMS (139 fb$^{-1}$) searched for di-boson resonances \cite{CMS:2021klu} ($\texttt{CMS:2021klu}$) , where the resonance has a mass range between 1 TeV to 4.5 TeV. The final states are those containing leptons and hadrons. In the models, we observe that all ruled instances are associated with production via VBF for which the upper limits range from 0.0086 pb (1 TeV) to 0.00016 pb (4.5 TeV), as can be seen in Figures \ref{fig:H3_t2} (T2), \ref{fig:H3_tx} (TX), and \ref{fig:H3_ty} (TY). Some points in each type, especially for $H_3$, might evade this constraint due to the breakdown of NWA, and are shown in black.

Finally for this class, the ATLAS detector (with 139 fb$^{-1}$) reported limits on heavy resonances decaying into a pair of gauge bosons in the semileptonic final states (i.e. one V decaying into leptons, while the other into hadrons) \cite{ATLAS:2020fry} ($\texttt{ATLAS:2020fry}$). The relevant mass range for the heavy scalar is between 300 and 5000 GeV. The search considered different production topologies, the one relevant to here is that of VBF, as selected by the $\mathtt{HB}$ criteria. The upper limit ranges from 0.604 pb to 0.00024. The effect of this search is seen as yellow points in the right panels of Figure \ref{fig:H3_t2} (T2), and Figures \ref{fig:H2_ty} and \ref{fig:H3_ty} (TY) where $m_H > 600$ GeV. Some points black points may pass this constraints due to being of large width, especially for TY ($H_2, H_3$) and T2 ($H_2, H_3$).

\subsubsection{Class: $pp \rightarrow H \rightarrow f \bar{f} $}
ATLAS and CMS searched for additional Higgs bosons decaying into fermions.  CMS carried out a dedicated search (138 fb$^{-1})$ in the $\tau \bar{\tau}$ final states \cite{CMS:2022goy} ($\texttt{CMS:2022goy}$). This search affects a mass range from 60 to 3500 GeV and considers production via ggH and bbH. The upper limits range from $\mathcal{O}(10)$ pb (60 GeV) to 0.003 pb (3.5 TeV). In our analysis, we find that this search affects the low mass range of $H_2$ and $H_3$, especially for TX. 

\subsection{Constraints on Effective Couplings}
To understand how the combined constraints affect the parameters of the four types, we delve into their effects on the effective couplings, focusing on $H_2$ and $H_3$. Not only do they determine the size of the branching ratios to different SM particles, but they also determine which production channel is dominant. Particularly, sizable couplings to up-type quarks accompanied with small coupling to vector bosons mean that ggH is larger than VBF, and vice versa. Also, since in T1, the effective couplings to fermions are equal, ggH is always larger than bbH. This is not the case for T2, for which bbH can be sizable, especially in the smaller mass range. 

\begin{figure}[htbp]
    \centering
    
    %--- T1 ---
    \begin{subfigure}{\textwidth}
        \centering
        \includegraphics[width=0.4\textwidth]{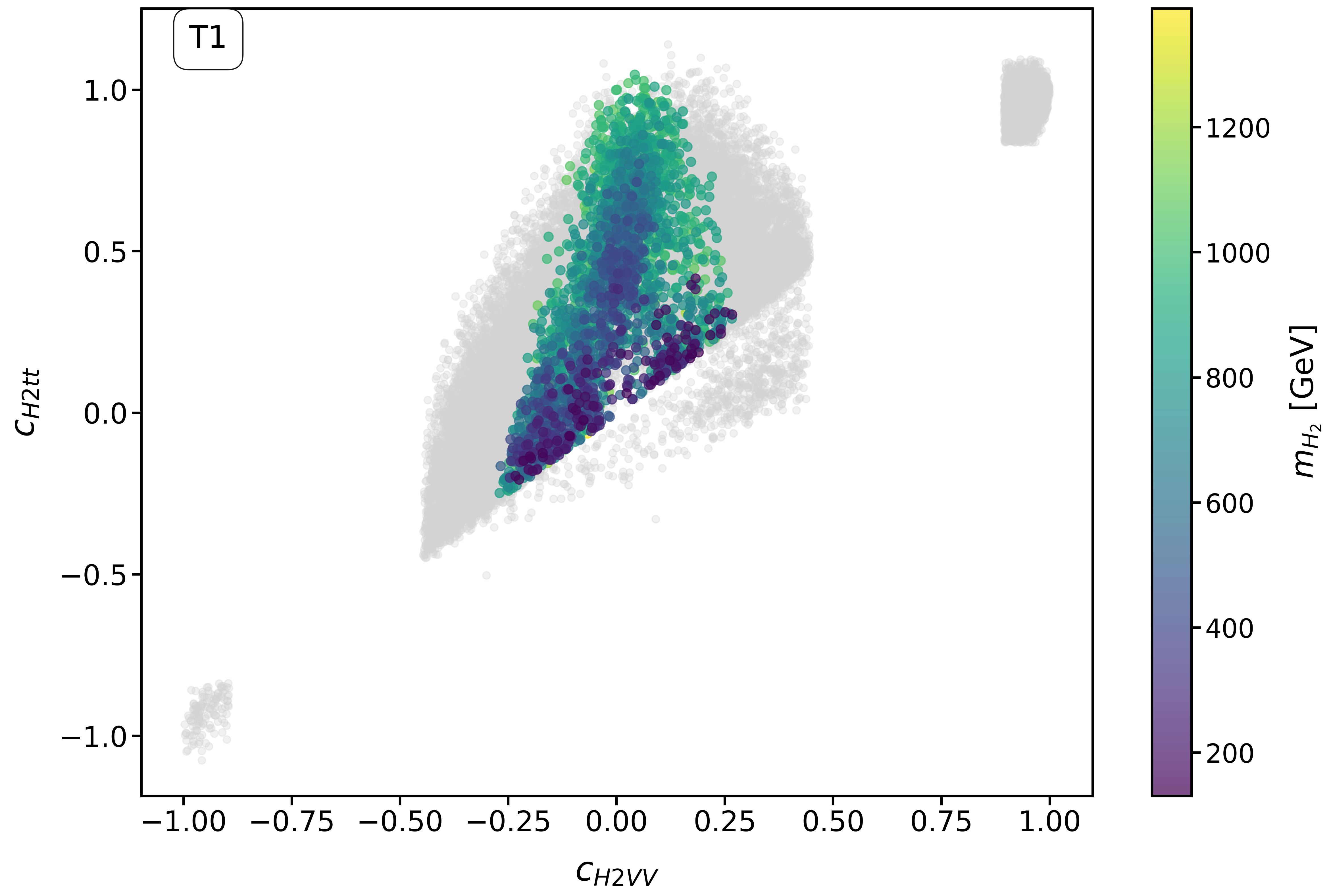}
        \label{fig:couplings_T1}
    \end{subfigure}
    \vspace{1em}
    
    %--- T2 ---
    \begin{subfigure}{\textwidth}
        \centering
        \includegraphics[width=0.7\textwidth]{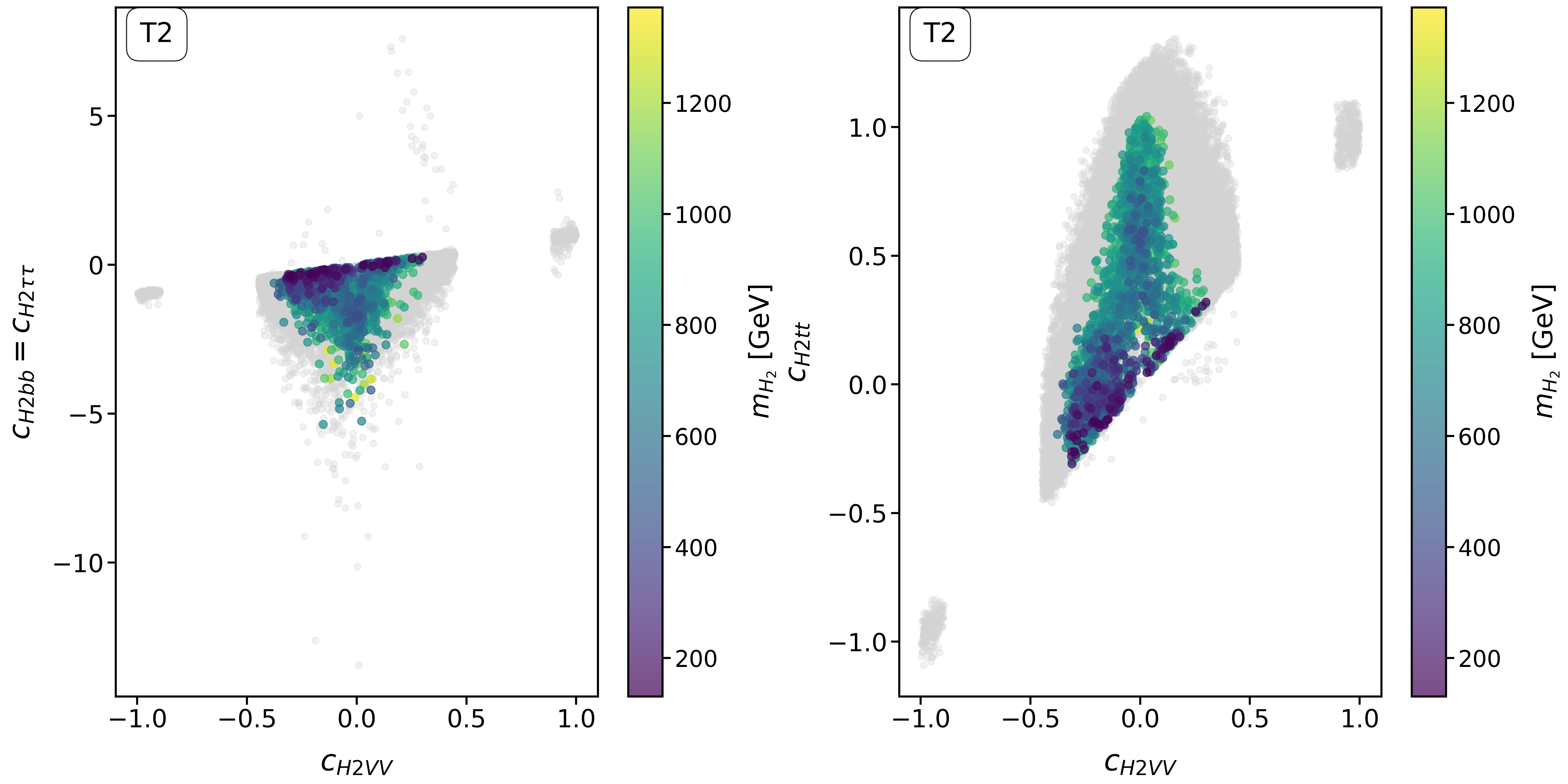}
        \label{fig:couplings_T2}
    \end{subfigure}
    \vspace{1em}
    
    %--- T3 ---
    \begin{subfigure}{\textwidth}
        \centering
        \includegraphics[width=0.7\textwidth]{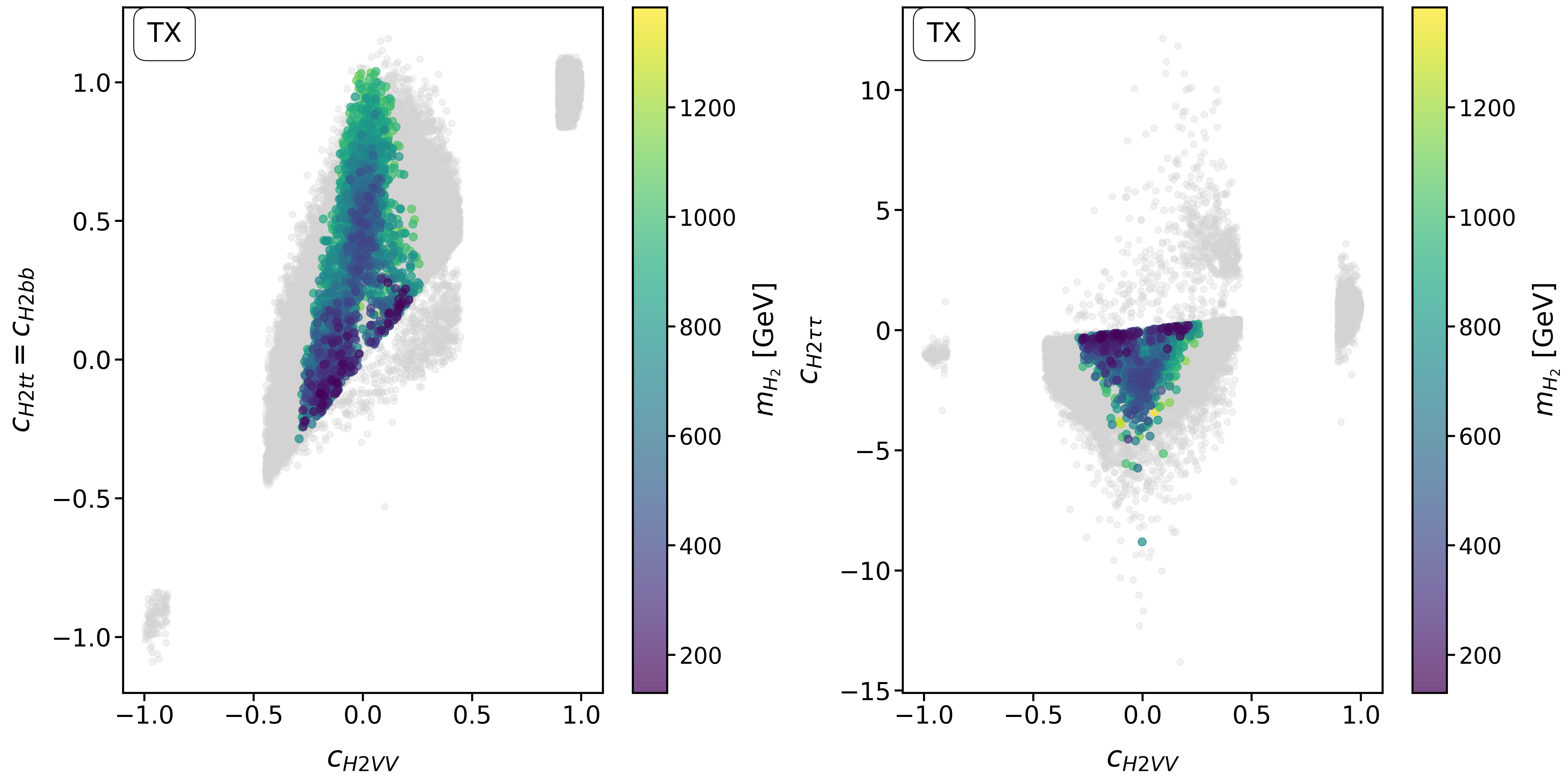}
        \label{fig:couplings_T3}
    \end{subfigure}
    \vspace{1em}
    
    %--- T4 ---
    \begin{subfigure}{\textwidth}
        \centering
        \includegraphics[width=0.7\textwidth]{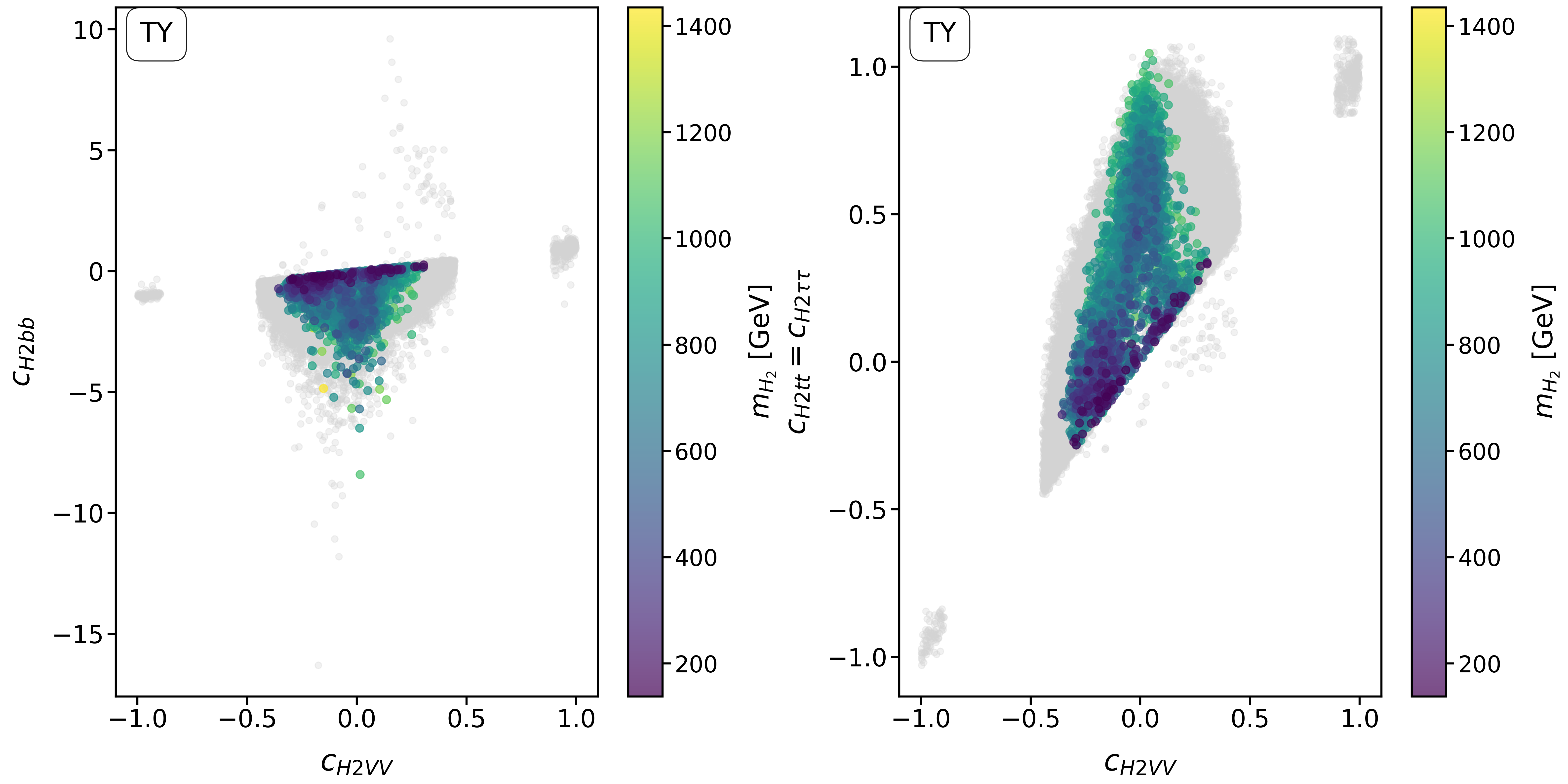}
        \label{fig:couplings_T4}
    \end{subfigure}
    
    \caption{Constraints on the effective couplings of the additional 
      CP-even Higgs boson $H_2$ to gauge bosons and fermions.}
    \label{fig:couplings_H2}
\end{figure}

\begin{figure}[htbp]
    \centering
    
    %--- T1 ---
    \begin{subfigure}{\textwidth}
        \centering
        \includegraphics[width=0.4\textwidth]{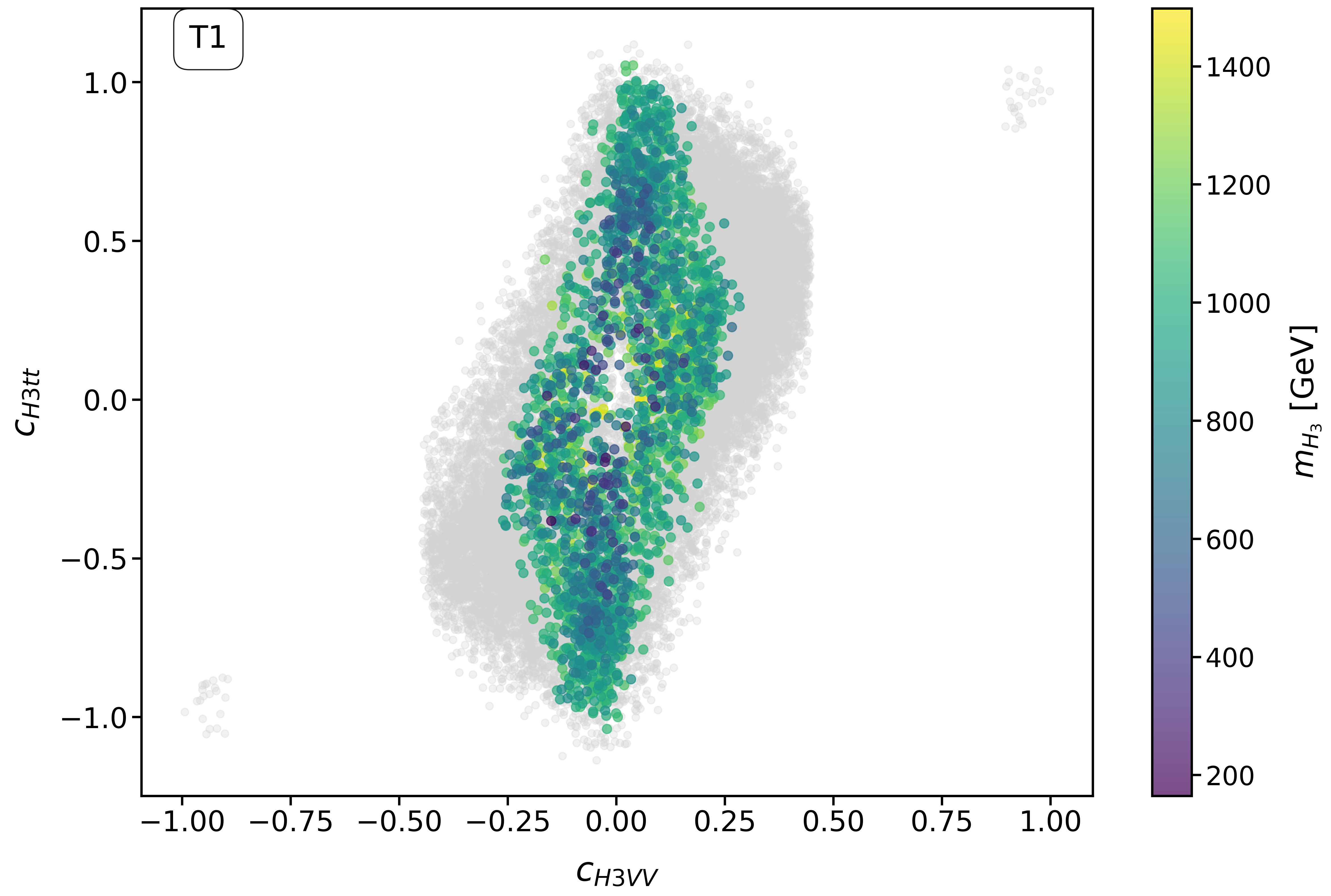}
        \label{fig:h3couplings_T1}
    \end{subfigure}
    \vspace{1em}
    
    %--- T2 ---
    \begin{subfigure}{\textwidth}
        \centering
        \includegraphics[width=0.7\textwidth]{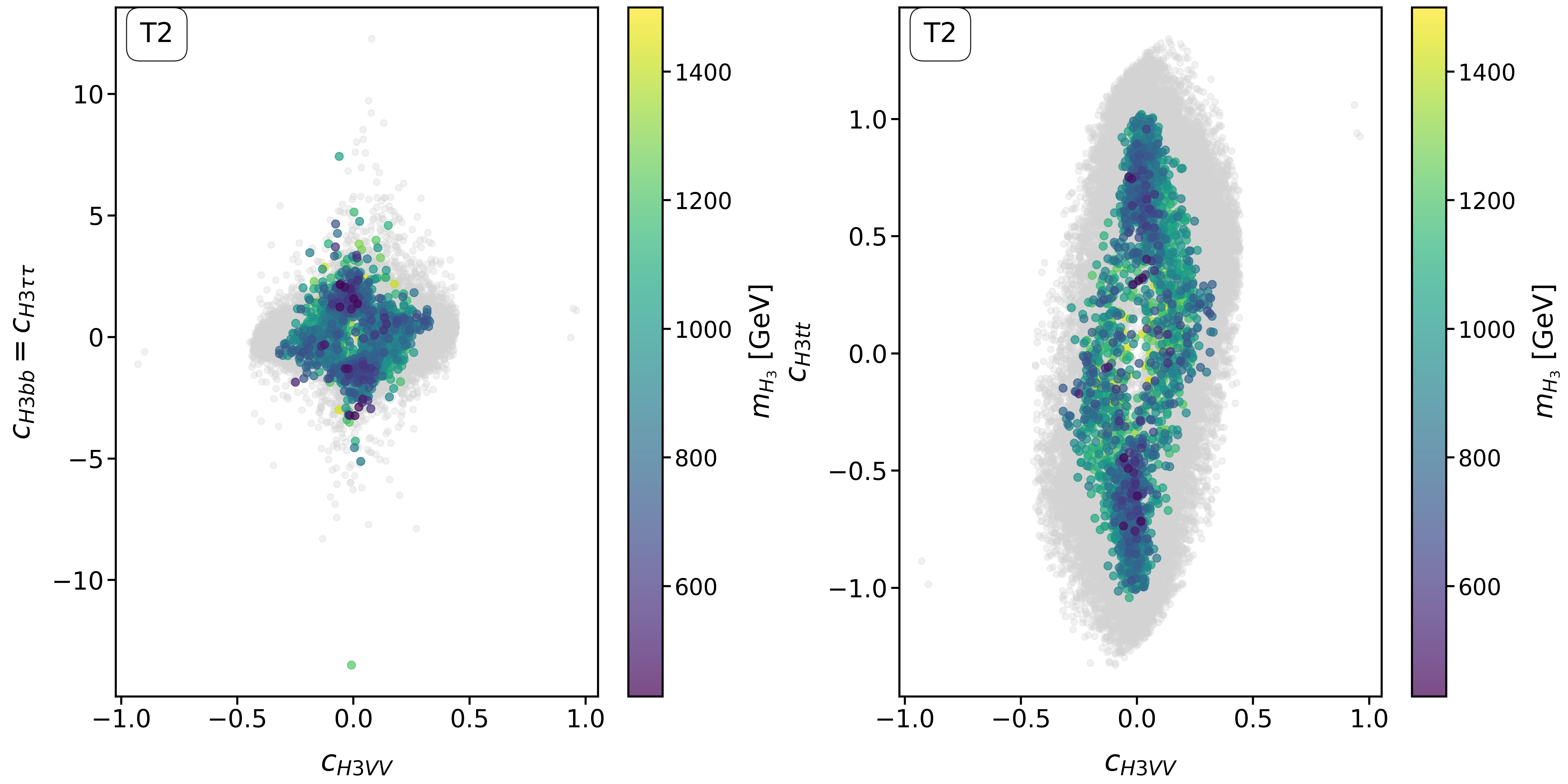}
        \label{fig:h3couplings_T2}
    \end{subfigure}
    \vspace{1em}
    
    %--- T3 ---
    \begin{subfigure}{\textwidth}
        \centering
        \includegraphics[width=0.7\textwidth]{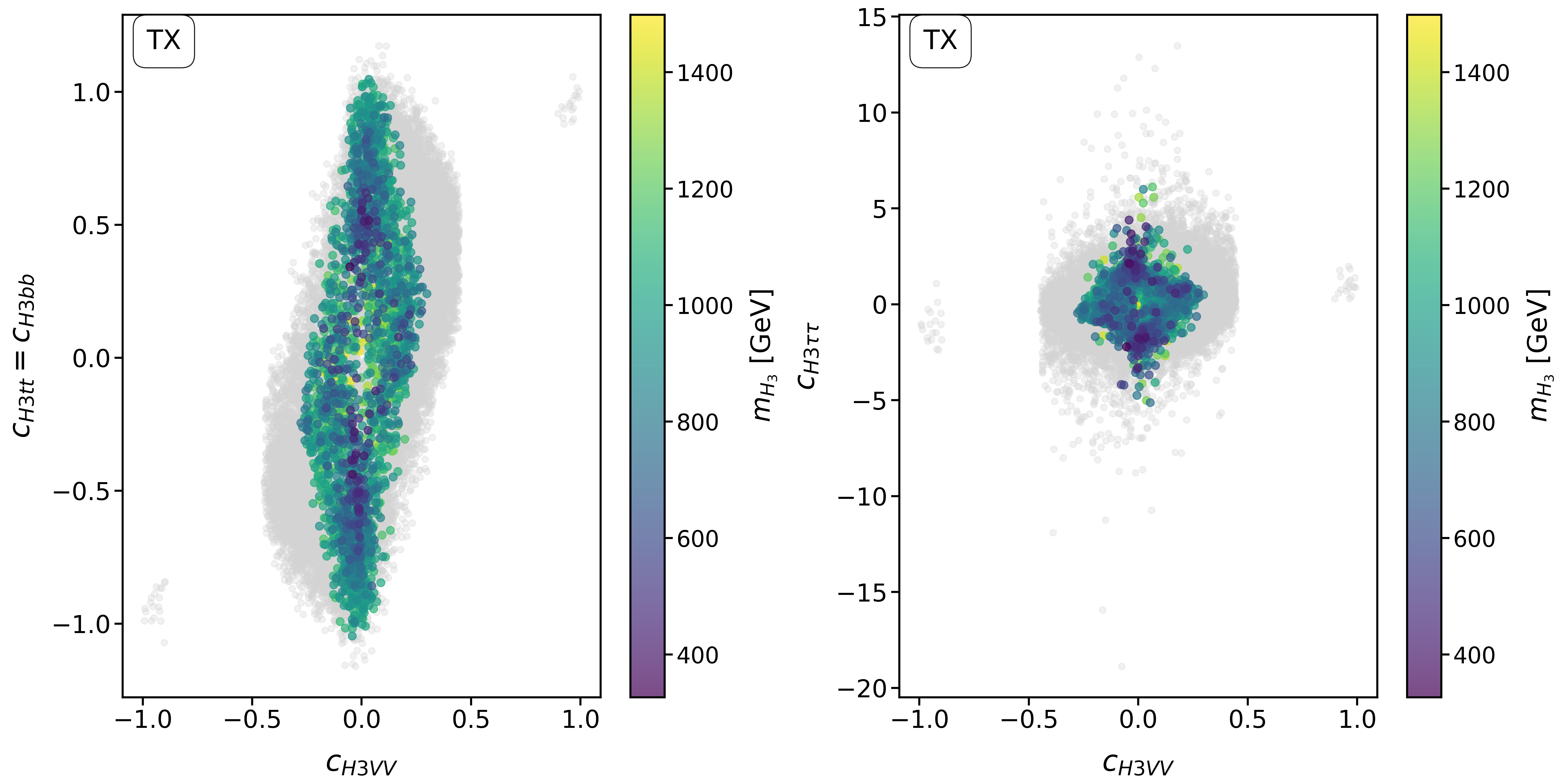}
        \label{fig:h3couplings_T3}
    \end{subfigure}
    \vspace{1em}
    
    %--- T4 ---
    \begin{subfigure}{\textwidth}
        \centering
        \includegraphics[width=0.7\textwidth]{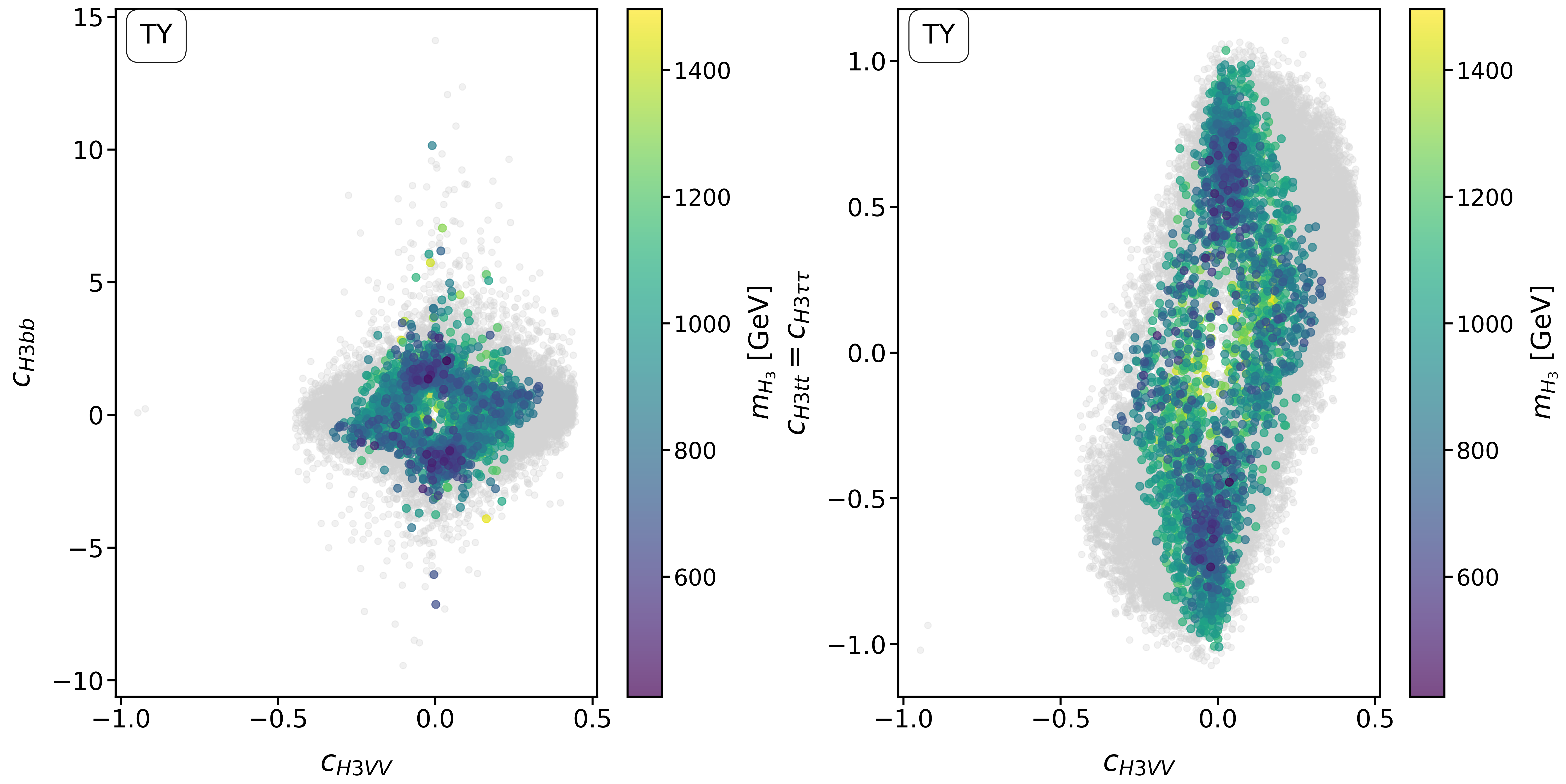}
        \label{fig:h3couplings_T4}
    \end{subfigure}
    
    \caption{Constraints on the effective couplings of the additional CP-even Higgs boson $H_3$ to gauge bosons and fermions.}
    \label{fig:couplings_H3}
\end{figure}

Restricting ourselves to the case where all additional Higgs bosons are narrow (including $A$ and $H^{\pm}$), Figures \ref{fig:couplings_H2} and \ref{fig:couplings_H3} show the effective couplings of $H_2$ and $H_3$ where the color code represents the mass of the additional Higgs boson sorted in descending order such that the smaller values are plotted on top of heavier ones. We can see in Figure \ref{fig:couplings_H2} that the smaller range of the mass is associated with small effective couplings, in general. We can further see that the effective couplings to bottom quarks (T2, TY) and $\tau$ leptons (T2, TX) are closer to zero for the small mass range. As the mass increases, these effective couplings open up and can be significantly larger than the effective couplings to gauge bosons. This constraint is not seen for the $H_3$ depicted in Figure \ref{fig:couplings_H3}, where the small mass range can still have effective couplings to fermions much larger than those to gauge bosons. 

The mass distribution shows distinctive patterns across the parameter space, with lower masses (400-600 GeV) concentrating in regions of smaller couplings for $H_2$, while $H_3$ exhibits a more spread distribution, forming characteristic triangular patterns in T1 that become increasingly asymmetric in other types.

In each figure, the subplots for T1, T2, TX, and TY are arranged to highlight how different types cluster in different coupling regions. For example, in T2 one observes that $c_{H_2bb}$ can become as negative as around $-5$, indicating a much larger deviation from SM-like behavior compared to T1. Similarly, in TY, the range of $c_{H_2bb}$ reaches around $-8.414$, and in TX the range of $c_{H_2 \tau \tau}$ can reach approximately $-8.811$. Meanwhile, the couplings to vector bosons (e.g.\ $c_{H_2VV}$) remain in a tighter range, rarely below $-0.376$ or above $0.305$ in all four types. The same pattern is even more pronounced for $H_3$, where in T2 the effective coupling to the bottom quark coupling can drop to $-13.499$ or rise above $7$, and in TY it can reach about $10.148$. These wide spreads for fermion couplings contrast with the relatively narrower window for $c_{H_3VV}$, spanning around $[-0.324, 0.331]$.

These coupling patterns have direct implications for the dominant production mechanisms, with the large variations in fermion couplings particularly affecting the interplay between gluon fusion and bottom-quark associated production across different mass ranges.

Finally, Table \ref{tab:coupling_ranges} presents the ranges allowed for effective couplings. One can see that the effective couplings to bottom quarks and tau leptons can significantly deviate to larger values compared with SM-like Higgs with the same mass. On the other hand, the coupling to top quarks can be smaller or close to SM-like Higgs of the same mass, while the effective couplings to vector bosons are always smaller than SM-like Higgs of the same mass. 
As indicated earlier, T2 and TY can reach substantial negative values for $c_{H_i bb}$, down to $-5.359$ and $-8.414$ for $H_2$, and even lower for $H_3$. The effective coupling with tau has sizable variations $[-8.811, 0.255]$ in TX for $H_2$, and $[-5.124, 6.119]$ in TX for $H_3$. In contrast, $c_{H_iVV}$ remains within comparatively narrow bounds $[-0.376,0.305]$ for $H_2$ in T2, and $[-0.324,0.331]$ for $H_3$ in TY.

\begin{table}[H]
    \caption{Ranges of allowed effective couplings for $H_2$ and $H_3$.}
    \label{tab:coupling_ranges}
    %\centering
    \begin{tabularx}{\textwidth}{llcccc}
        \hline\hline
        Particle & Type & $c_{H_iVV}$ & $c_{H_iuu}$ & $c_{H_idd}$ & $c_{H_ill}$ \\
        \hline
        \multirow{4}{*}{$H_2$} 
        & T1  & $[-0.270, 0.267]$ & $[-0.248, 1.046]$ & $c_{H_2dd} = c_{H_2uu}$ & $c_{H_2ll} = c_{H_2uu}$ \\[1ex]
        & T2  & $[-0.376, 0.300]$ & $[-0.309, 1.042]$ & $[-5.359, 0.260]$ & $c_{H_2ll} = c_{H_2dd}$ \\[1ex]
        & TX  & $[-0.291, 0.259]$ & $[-0.285, 1.039]$ & $c_{H_2dd} = c_{H_2uu}$ & $[-8.811, 0.255]$ \\[1ex]
        & TY  & $[-0.357, 0.305]$ & $[-0.282, 1.045]$ & $[-8.414, 0.258]$ & $c_{H_2ll} = c_{H_2uu}$ \\[1ex]
        \hline
        \multirow{4}{*}{$H_3$} 
        & T1  & $[-0.261, 0.284]$ & $[-1.037, 1.053]$ & $c_{H_3dd} = c_{H_3uu}$ & $c_{H_3ll} = c_{H_3uu}$ \\[1ex]
        & T2  & $[-0.318, 0.329]$ & $[-1.042, 1.019]$ & $[-13.499, 7.429]$ & $c_{H_3ll} = c_{H_3dd}$ \\[1ex]
        & TX  & $[-0.278, 0.298]$ & $[-1.047, 1.047]$ & $c_{H_3dd} = c_{H_3uu}$ & $[-5.124, 6.119]$ \\[1ex]
        & TY  & $[-0.324, 0.331]$ & $[-1.010, 1.036]$ & $[-7.138, 10.148]$ & $c_{H_3ll} = c_{H_3uu}$ \\[1ex]
        \hline\hline
    \end{tabularx}
\end{table}

\subsection{Constraints on Pseudoscalar and Charged Higgs Bosons}
\begin{figure}[H]
        \includegraphics[width=\textwidth]{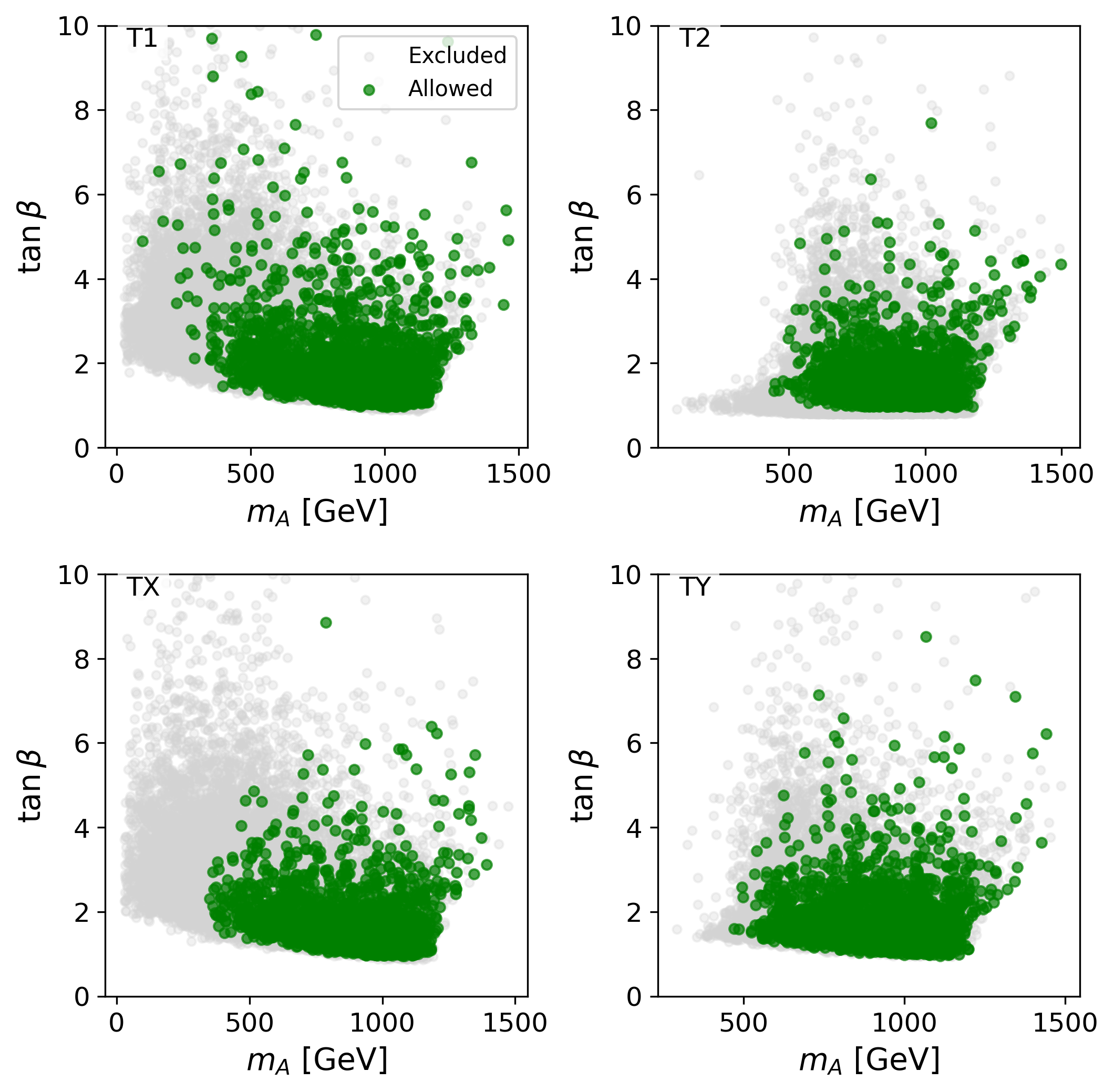}
        \caption{Allowed (green) and excluded (grey) points in the $m_A$-$\tan\beta$ plane.}
        \label{fig:mA_tb}
\end{figure}

\begin{figure}[H]
   \includegraphics[width=\textwidth]{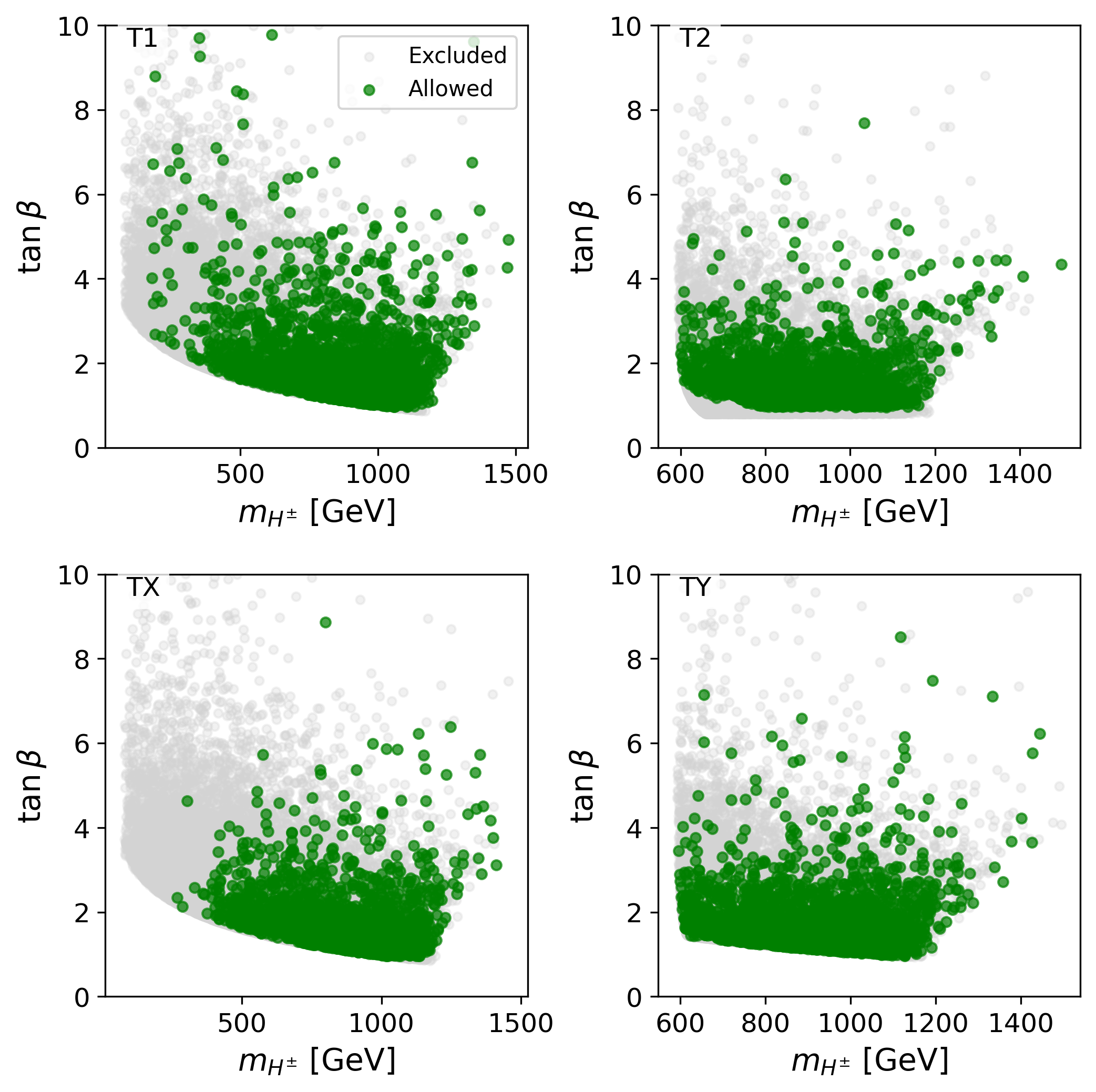}
        \caption{Allowed (green) and excluded (grey) points in the $m_{H^\pm}$-$\tan\beta$ plane.}
        \label{fig:mX_tb}
\end{figure}

As mentioned in the Introduction, the pseudoscalar $A$, and the charged Higgs pair $H^{\pm}$ have the same structure as in 2HDM. However, these can be indirectly affected by the overall constraints on the N2HDM. Since the effective couplings of $A$ and $H^{\pm}$ depend on $\tan \beta$, Figures \ref{fig:mA_tb} and \ref{fig:mX_tb} show the allowed parameter spaces in the $m_{A/H^{\pm}}$-$\tan\beta$ planes for the four types. 

The distributions in the $m_A$-$\tan\beta$ plane exhibit distinctive patterns across all types. T1 shows the widest mass range, allowing pseudoscalar masses as low as 97.8 GeV and extending up to about 1460 GeV, with $\tan\beta$ values reaching up to approximately 12. T2 shows a more constrained parameter space, with $m_A$ starting at around 446 GeV, and most allowed points are concentrated at lower values of $\tan\beta$. TX shows an intermediate mass range starting from 350 GeV, while TY has a higher mass threshold around 470 GeV. The charged Higgs masses follow similar patterns, with T1 showing the broadest range starting from 179.7 GeV, while T2 and TY have higher thresholds around 600 GeV. TX allows for relatively lighter charged Higgs with masses starting from 268 GeV. In all types, the parameter space becomes increasingly sparse at higher $\tan\beta$ values, with the densest populations observed below $\tan\beta \approx 4$.

In T1, the pseudoscalar $A$ with a mass range between 225 and 1000 GeV, is mainly affected by CMS search for $A$ decaying into Z ($Z\rightarrow ll/\nu \nu$) and an SM-like Higgs boson ($h \rightarrow bb$) \cite{CMS:2019qcx}. As for T2, we observe that some points are ruled out because $A$ is inconsistent with measurements from ATLAS \cite{ATLAS:2023szc}, where it decays into a heavy Higgs (decaying into top pairs) and a Z boson. This is relevant for mass ranges for A between 450 and 1200 GeV and heavy Higgs between 350 and 800 GeV. Additionally, and for TX, we observe that $A$ is affected by the likelihood analysis presented by the CMS search for additional Higgs bosons decaying into a pair of $\tau$'s \cite{CMS:2022goy} for a mass between 160 GeV and 3500 GeV. We also find that for TY, $A$ is mostly affected by ATLAS searches for a $A$ decaying into heavy Higgs and Z bosons \cite{ATLAS:2023szc, ATLAS:2020gxx}. The former is for $m_A > 800$ GeV and $m_H > 300$ GeV, while the latter is for $230 \leq m_A \ (\text{GeV}) \leq 800$ and $130 \leq m_{H} \ (\text{GeV}) \leq 700$. 

On the other hand, for all types, the charged Higgs boson is affected by results from ATLAS \cite{ATLAS:2021upq} searching for $pp \rightarrow tbH^+ \rightarrow tbtb$, where $200 \leq m_{H^{\pm}} \ (\text{GeV}) \leq 2000$ \footnote{$\mathtt{HT}$ extends the range to start from $m_{H^{\pm}} \geq 145 $ GeV.}, in final states consisting of jets and one electron or muon. For a mass range smaller than 145 GeV, which only relevant for T1 and TX, the constraint on the charged Higgs is mainly set, based on $\mathtt{HB}$ selection, by ATLAS search for charged Higgs decaying into $\tau \nu_\tau$ \cite{ATLAS:2018gfm}, where the charged Higgs is produced in decays of the top quark.

\subsection{Prospects for future searches}
The LHC is expected to be upgraded to the High-Luminosity LHC (HL-LHC) \cite{ZurbanoFernandez:2020cco} by 2030, allowing for the allocation of 3000 fb$^{-1}$ of data during its operation. This will enable more precise measurements of the properties of the SM-like Higgs boson and expand the discovery reach for additional scalars. In particular, the projected precision for measuring the main Higgs production channels ranges from 1.6$\%$ (ggH) to 5.7$\%$ (WH). Meanwhile, dominant decay modes are expected to be probed with precisions of 2.6$\%$ ($\gamma \gamma$), about 2.9$\%$ (ZZ, $W^+ W^-$, $\tau^+ \tau^-$), and 4.4$\%$ ($b \bar{b}$). Rare decays to $\mu^+ \mu^-$ and $Z\gamma$ are expected to be observed, but with larger uncertainties. In terms of coupling modifiers, the projected uncertainties are also at the a few percent level \cite{Cepeda:2019klc, ATLAS:2022hsp}. These measurements will have important implications for the parameter spaces of BSM extensions such as the N2HDM.

Furthermore, searches for BSM Higgs bosons will considerably improve, extending the reach in probing mass ranges and couplings by up to 50$\%$ \cite{CidVidal:2018eel, Cepeda:2019klc, ATLAS:2022hsp}. For instance, limits on the decay of a heavy scalar resonance into a pair of Z bosons are anticipated to improve by a factor of ten. Processes such as $A \rightarrow ZH$ and $H \rightarrow ZA$ will become important probes, especially in regions with substantial mass splitting, and decays to a pair of $\tau$ leptons will provide complementary channels. Parameter regions away from the alignment limit can be tested via $pp \rightarrow A \rightarrow Zh \rightarrow \ell \ell b \bar{b}$, which is expected to gain sensitivity in further upgrades at $\sqrt{s}$ between 14 and 27 TeV (The High Energy LHC). The overall implications on the different types of the N2HDM will be significant in terms of restricting the allowed SM-like Higgs couplings, the mixing angles, the singlet component, as well as the allowed mass range and properties of the additinoal Higgs boson. A detailed analysis of such effects will be the subject of a future work.   

\section{Conclusions} \label{con} All in all, we have considered the broken phase of the N2HDM where the two Higgs doublets ($\Phi_1$ and $\Phi_2$) and the singlet ($\Phi_S$) acquire VEVs. The model admits two discrete symmetries, one of which is spontaneously broken by the singlet VEV, while the other is softly broken by the $m^{2}_{12}$ parameter and extends to the Yukawa sector. This brings about four types that encode the different possibilities of Yukawa couplings between the Higgs doublets and SM fermions (i.e. Type 1, Type 2, Type X, and Type Y). The parameter spaces have been subjected to limits from theory and observations. The model was interfaced with the latest Higgs data repositories of the public code $\mathtt{HiggsTools}$ and its subpackages.

Performing a statistical $\chi^2$ analysis using $\mathtt{HS}$, we identified the best-fit point for each type. We found that T1 is slightly shifted from the alignment limit with $\tan\beta$ being as small as 0.96, while the best-fit points of the other types reside within the alignment limit with moderate values of $\tan\beta \sim 5$. We analyzed the specific CMS and ATLAS measurements affecting the neighboring regions of the best-fit points, leading to deviations from $\chi^2_{\text{min}}$. Furthermore, we determined the effects of the constraints on the parameters of the model, namely, the singlet component of the SM-like Higgs, and the mixing angles.

We have also analyzed the bounds on the additional Higgs bosons using $\mathtt{HB}$, particularly from searches for resonance production of a pair of bosons or fermions via an additional Higgs. Additionally, we have shown the effects of the constraints on the effective couplings of the additional Higgs bosons, and their masses, including the pseudoscalar and charged Higgs pair.

In particular, we observe that the SM-like Higgs signals were mainly constrained by searches involving $h \rightarrow VV, \gamma\gamma, b\bar{b}, \tau^+\tau^-$. The mixing angles $\alpha_1$ and $\alpha_2$ are subject to the ranges summarized in Table~\ref{tabalpha}, while the singlet component of SM-like Higgs is below $10\%$ in T1 and TX, below $15\%$ in T2, and can reach up to $18\%$ in TY. Concerning the additional Higgs bosons, the most restrictive classes of measurements are those searching for heavy resonances decaying into $hh$ and $VV$.

We find that the constraints on the effective couplings to vector bosons, $c_{H_iVV}$, are quite stringent in all types, requiring them to lie in comparatively narrow ranges, such as $-0.376 \leq c_{H_2VV} \leq 0.305$ in T2 or $-0.324 \leq c_{H_3VV} \leq 0.331$ in TY. Meanwhile, the effective couplings to fermions can be significantly larger than the SM, for instance, in T2 and TY the bottom-quark coupling $c_{H_3bb}$ can range down to about $-13.5$ and up to about $10.1$, respectively. Regarding the pseudoscalar $A$, the allowed mass range in T1 spans from about 98 GeV to about 1460 GeV, in T2 it starts around 446 GeV, in TX around 350 GeV, and in TY around 470 GeV. The corresponding $\tan\beta$ values can be as small as around 0.8 or as large as 12, depending on the type.

Finally, the overall constraints from the Higgs data show the vital role played by searches for additional Higgs bosons. It is important to keep up-to-date with results from recent and future LHC runs. These are expected to be systematically included in updates by the $\texttt{HT}$ group, and our interface/analysis code can be readily utilized to inspect any further effects on the four types of the N2HDM.

\section*{Acknowledgement}
I thank Henning Bahl for helpful correspondence regarding HiggsTools. This work is supported by King Saud University.

\section*{Appendix A}
\subsection*{A-1: Effective Couplings in Type 1}

For Type 1, the effective couplings are:
\begin{align}
c_{H_1 f\bar{f}} &= \frac{\sin \alpha_1 \cos \alpha_2}{\sin \beta}, \\
c_{H_1 VV} &= \cos \alpha_1 \cos \alpha_2 \cos \beta + \sin \alpha_1 \cos \alpha_2 \sin \beta.
\end{align}

For the second scalar $H_2$:
\begin{align}
c_{H_2 f\bar{f}} &= \frac{\cos \alpha_1 \cos \alpha_3 - \sin \alpha_1 \sin \alpha_2 \sin \alpha_3}{\sin \beta}, \\
c_{H_2 VV} &= -\left( \cos \alpha_1 \sin \alpha_2 \sin \alpha_3 + \sin \alpha_1 \cos \alpha_3 \right) \cos \beta \\
&\quad + \left( \cos \alpha_1 \cos \alpha_3 - \sin \alpha_1 \sin \alpha_2 \sin \alpha_3 \right) \sin \beta.
\end{align}

For the third scalar $H_3$:
\begin{align}
c_{H_3 f\bar{f}} &= \frac{-\cos \alpha_1 \sin \alpha_3 - \sin \alpha_1 \sin \alpha_2 \cos \alpha_3}{\sin \beta}, \\
c_{H_3 VV} &= \left( -\cos \alpha_1 \sin \alpha_2 \cos \alpha_3 + \sin \alpha_1 \sin \alpha_3 \right) \cos \beta \\
&\quad + \left( -\cos \alpha_1 \sin \alpha_3 - \sin \alpha_1 \sin \alpha_2 \cos \alpha_3 \right) \sin \beta.
\end{align}

For the pseudoscalar $A$:
\begin{align}
c_{A f\bar{f}} &= \frac{1}{\tan \beta}.
\end{align}

\subsection*{A-2: Effective Couplings in Type 2}

For Type 2, the effective couplings are:
\begin{align}
c_{H_1 tt} &= \frac{\sin \alpha_1 \cos \alpha_2}{\sin \beta}, \\
c_{H_1 bb} &= c_{H_1 \tau \bar{\tau}} = \frac{\cos \alpha_1 \cos \alpha_2}{\cos \beta}, \\
c_{H_1 VV} &= \cos \alpha_1 \cos \alpha_2 \cos \beta + \sin \alpha_1 \cos \alpha_2 \sin \beta.
\end{align}

For the second scalar $H_2$:
\begin{align}
c_{H_2 tt} &= \frac{\cos \alpha_1 \cos \alpha_3 - \sin \alpha_1 \sin \alpha_2 \sin \alpha_3}{\sin \beta}, \\
c_{H_2 bb} &= c_{H_2 \tau \bar{\tau}} = \frac{-\cos \alpha_1 \sin \alpha_2 \sin \alpha_3 - \sin \alpha_1 \cos \alpha_3}{\cos \beta}, \\
c_{H_2 VV} &= -\left( \cos \alpha_1 \sin \alpha_2 \sin \alpha_3 + \sin \alpha_1 \cos \alpha_3 \right) \cos \beta \\
&\quad + \left( \cos \alpha_1 \cos \alpha_3 - \sin \alpha_1 \sin \alpha_2 \sin \alpha_3 \right) \sin \beta.
\end{align}

For the third scalar $H_3$:
\begin{align}
c_{H_3 tt} &= \frac{-\cos \alpha_1 \sin \alpha_3 - \sin \alpha_1 \sin \alpha_2 \cos \alpha_3}{\sin \beta}, \\
c_{H_3 bb} &= c_{H_3 \tau \bar{\tau}} = \frac{-\cos \alpha_1 \sin \alpha_2 \cos \alpha_3 + \sin \alpha_1 \sin \alpha_3}{\cos \beta}, \\
c_{H_3 VV} &= \left( -\cos \alpha_1 \sin \alpha_2 \cos \alpha_3 + \sin \alpha_1 \sin \alpha_3 \right) \cos \beta \\
&\quad + \left( -\cos \alpha_1 \sin \alpha_3 - \sin \alpha_1 \sin \alpha_2 \cos \alpha_3 \right) \sin \beta.
\end{align}

For the pseudoscalar $A$:
\begin{align}
c_{A tt} &= \frac{1}{\tan \beta}, \\
c_{A bb} &= \tan \beta.
\end{align}

\subsection*{A-3: Effective Couplings in LS}

For Type 3, the effective couplings are:
\begin{align}
c_{H_1 tt} &= c_{H_1 bb} = \frac{\sin \alpha_1 \cos \alpha_2}{\sin \beta}, \\
c_{H_1 \tau \bar{\tau}} &= \frac{\cos \alpha_1 \cos \alpha_2}{\cos \beta}, \\
c_{H_1 VV} &= \cos \alpha_1 \cos \alpha_2 \cos \beta + \sin \alpha_1 \cos \alpha_2 \sin \beta.
\end{align}

For the second scalar $H_2$:
\begin{align}
c_{H_2 tt} &= c_{H_2 bb} = \frac{\cos \alpha_1 \cos \alpha_3 - \sin \alpha_1 \sin \alpha_2 \sin \alpha_3}{\sin \beta}, \\
c_{H_2 \tau \bar{\tau}} &= \frac{-\cos \alpha_1 \sin \alpha_2 \sin \alpha_3 - \sin \alpha_1 \cos \alpha_3}{\cos \beta}, \\
c_{H_2 VV} &= -\left( \cos \alpha_1 \sin \alpha_2 \sin \alpha_3 + \sin \alpha_1 \cos \alpha_3 \right) \cos \beta \\
&\quad + \left( \cos \alpha_1 \cos \alpha_3 - \sin \alpha_1 \sin \alpha_2 \sin \alpha_3 \right) \sin \beta.
\end{align}

For the third scalar $H_3$:
\begin{align}
c_{H_3 tt} &= c_{H_3 bb} = \frac{-\cos \alpha_1 \sin \alpha_3 - \sin \alpha_1 \sin \alpha_2 \cos \alpha_3}{\sin \beta}, \\
c_{H_3 \tau \bar{\tau}} &= \frac{-\cos \alpha_1 \sin \alpha_2 \cos \alpha_3 + \sin \alpha_1 \sin \alpha_3}{\cos \beta}, \\
c_{H_3 VV} &= \left( -\cos \alpha_1 \sin \alpha_2 \cos \alpha_3 + \sin \alpha_1 \sin \alpha_3 \right) \cos \beta \\
&\quad + \left( -\cos \alpha_1 \sin \alpha_3 - \sin \alpha_1 \sin \alpha_2 \cos \alpha_3 \right) \sin \beta.
\end{align}

For the pseudoscalar $A$:
\begin{align}
c_{A f\bar{f}} &= \frac{1}{\tan \beta}.
\end{align}

\subsection*{A-4: Effective Couplings in FL}

For Type 4, the effective couplings are:
\begin{align}
c_{H_1 tt} &= c_{H_1 \tau \bar{\tau}} = \frac{\sin \alpha_1 \cos \alpha_2}{\sin \beta}, \\
c_{H_1 bb} &= \frac{\cos \alpha_1 \cos \alpha_2}{\cos \beta}, \\
c_{H_1 VV} &= \cos \alpha_1 \cos \alpha_2 \cos \beta + \sin \alpha_1 \cos \alpha_2 \sin \beta.
\end{align}

For the second scalar $H_2$:
\begin{align}
c_{H_2 tt} &= c_{H_2 \tau \bar{\tau}} = \frac{\cos \alpha_1 \cos \alpha_3 - \sin \alpha_1 \sin \alpha_2 \sin \alpha_3}{\sin \beta}, \\
c_{H_2 bb} &= \frac{-\cos \alpha_1 \sin \alpha_2 \sin \alpha_3 - \sin \alpha_1 \cos \alpha_3}{\cos \beta}, \\
c_{H_2 VV} &= -\left( \cos \alpha_1 \sin \alpha_2 \sin \alpha_3 + \sin \alpha_1 \cos \alpha_3 \right) \cos \beta \\
&\quad + \left( \cos \alpha_1 \cos \alpha_3 - \sin \alpha_1 \sin \alpha_2 \sin \alpha_3 \right) \sin \beta.
\end{align}

For the third scalar $H_3$:
\begin{align}
c_{H_3 tt} &= c_{H_3 \tau \bar{\tau}} = \frac{-\cos \alpha_1 \sin \alpha_3 - \sin \alpha_1 \sin \alpha_2 \cos \alpha_3}{\sin \beta}, \\
c_{H_3 bb} &= \frac{-\cos \alpha_1 \sin \alpha_3 - \sin \alpha_1 \sin \alpha_2 \cos \alpha_3}{\sin \beta}, \\
c_{H_3 VV} &= \left( -\cos \alpha_1 \sin \alpha_2 \cos \alpha_3 + \sin \alpha_1 \sin \alpha_3 \right) \cos \beta \\
&\quad + \left( -\cos \alpha_1 \sin \alpha_3 - \sin \alpha_1 \sin \alpha_2 \cos \alpha_3 \right) \sin \beta.
\end{align}

For the pseudoscalar $A$:
\begin{align}
c_{A tt} &= \frac{1}{\tan \beta}, \\
c_{A bb} &= \tan \beta.
\end{align}

%\bibliographystyle{unsrt}
%\bibliography{N2HDMHT}

\end{document}